\documentclass[fleqn,usenatbib,useAMS]{mnras}

\usepackage[T1]{fontenc}
\usepackage{graphicx}
\DeclareGraphicsExtensions{.pdf,.png}
\usepackage[dvipsnames]{xcolor}
\usepackage{amsmath}
\usepackage{amssymb}
\usepackage{comment}
\usepackage[normalem]{ulem}
\usepackage{footnote}
\usepackage{dsfont}
\usepackage{enumitem}
\usepackage{bbold}
\usepackage{bm}
\usepackage{cprotect} 

\DeclareFontFamily{OT1}{pzc}{}
\DeclareFontShape{OT1}{pzc}{m}{it}{<-> s * [1.10] pzcmi7t}{}
\DeclareMathAlphabet{\mathpzc}{OT1}{pzc}{m}{it}

\definecolor{blue}{rgb}{0,0,1}
\definecolor{darkergreen}{rgb}{0,0.5,0}

\defcitealias{Planck2018PNG}{Planck Collaboration (2019)}

\newcommand{\eg}{e.g$.$ }

\newcommand{\egnospace}{e.g$.$}

\newcommand{\ie}{i.e$.$ }
\newcommand{\cf}{cf$.$ }
\newcommand{\cfnospace}{cf$.$}
\newcommand{\resp}{resp$.$ }
\newcommand{\wrt}{wrt$.$ }
\newcommand{\dd}{\ensuremath{\mathrm{d}}}

\newcommand{\figref}[1]{Figure \ref{fi:#1}}
\newcommand{\figrefnospace}[1]{Figure \ref{fi:#1}}
\newcommand{\Figref}[1]{Figure \ref{fi:#1}}

\newcommand{\tabrefnospace}[1]{Table \ref{tab:#1}}

\newcommand{\eqnref}[1]{Equation \ref{eq:#1}}
\newcommand{\eqnrefnospace}[1]{Equation \ref{eq:#1}}

\newcommand{\appref}[1]{Appendix \ref{app:#1}}
\newcommand{\apprefnospace}[1]{Appendix \ref{app:#1}}

\newcommand{\secref}[1]{Section \ref{sec:#1}}
\newcommand{\secrefnospace}[1]{Section \ref{sec:#1}}
\newcommand{\Secref}[1]{Section \ref{sec:#1}}

\allowdisplaybreaks

\title[The PDF perspective on the tracer-matter connection]{The PDF perspective on the tracer-matter connection: Lagrangian bias and non-Poissonian shot noise}
\author[]{\parbox{\linewidth}{Oliver Friedrich$^{1, 2}$, Anik Halder$^{3, 4}$, Aoife Boyle$^{5}$, Cora Uhlemann$^6$, Dylan Britt$^{7, 8}$, Sandrine Codis$^{9}$, Daniel Gruen$^{3, 8, 10}$, ChangHoon Hahn$^{11}$}
\vspace*{10pt}\\
$^1$ {Kavli Institute for Cosmology, University of Cambridge, CB3 0HA Cambridge, United Kingdom}\\
$^{2}$ {Churchill College, University of Cambridge, CB3 0DS Cambridge, United Kingdom}\\
$^{3}$ Universitäts-Sternwarte, Fakultät für Physik, Ludwig-Maximilians Universität München, Scheinerstr. 1, 81679 München, Germany\\
$^{4}$ Max Planck Institute for Extraterrestrial Physics, Giessenbachstrasse 1, 85748 Garching, Germany\\
$^{5}$ CNRS \& Sorbonne Universit\'e, UMR 7095, Institut d'Astrophysique de Paris, 75014, Paris, France\\
$^{6}$ {School of Mathematics, Statistics and Physics, Newcastle University, Herschel Building, NE1 7RU Newcastle-upon-Tyne, United Kingdom}\\
$^{7}$ Department of Physics, Stanford University, 382 Via Pueblo Mall, Stanford, CA 94305, USA\\
$^{8}$ Kavli Institute for Particle Astrophysics \& Cosmology, P. O. Box 2450, Stanford University, Stanford, CA 94305, USA\\
$^{9}$ AIM, CEA, CNRS, Université Paris-Saclay, Université Paris Diderot, Sorbonne Paris Cité, 91191 Gif-sur-Yvette, France\\
$^{10}$ SLAC National Accelerator Laboratory, Menlo Park, CA 94025, USA\\
$^{11}$ Department of Astrophysical Sciences, Princeton University, Peyton Hall, Princeton NJ 08544, USA\\
}

\begin{document}

\maketitle

\begin{abstract}
We study the connection of matter density and its tracers from the PDF perspective. One aspect of this connection is the conditional expectation value $\langle \delta_{\mathrm{tracer}}|\delta_m\rangle$ when averaging both tracer and matter density over some scale. We present a new way to incorporate a Lagrangian bias expansion of this expectation value into standard frameworks for modelling the PDF of density fluctuations and counts-in-cells statistics. Using N-body simulations and mock galaxy catalogs we confirm the accuracy of this expansion and compare it to the more commonly used Eulerian parametrization. For halos hosting typical luminous red galaxies, the Lagrangian model provides a significantly better description of $\langle \delta_{\mathrm{tracer}}|\delta_m\rangle$ at second order in perturbations. A second aspect of the matter-tracer connection is shot-noise, \ie the scatter of tracer density around $\langle \delta_{\mathrm{tracer}}|\delta_m\rangle$. It is well known that this noise can be significantly non-Poissonian and we validate the performance of a more general, two-parameter shot-noise model for different tracers and  simulations. Both parts of our analysis are meant to pave the way for forthcoming applications to survey data.
\end{abstract}

\section{Introduction}
\label{sec:introduction}

Studying the evolution of the cosmic density field with the help of galaxy positions is like trying to understand a mountain range from knowing the location of (some of) its mountain peaks. One can hardly hope to infer the full profile of the density field from (a subset of) its luminous tracers. But one can hope that statistical properties of the galaxy density field can be expressed as functions of corresponding statistical properties of the total matter density field. For example, in the case of 2-point statistics, one may assume that the galaxy clustering correlation function is just a multiple of the matter density correlation function (linear galaxy bias). In that case, any cosmological information contained in the shape of the matter density 2-point function can still be retrieved from the galaxy density 2-point function.

For such a program to be successful, one would optimally like to know the precise functional form that relates statistics of the matter density and galaxy density fields. And if there are unknown features in that functional form, then one would at least like to break down these features into a well defined set of unknown numbers that parametrize our ignorance. The earliest attempt at finding such a parametrization was made by \citet[][Kaiser bias]{Kaiser1984}, who found that at sufficiently large scales the 2-point function of collapsed objects (clusters as modeled by overdense regions) is indeed proportional to the 2-point function of the density field. At small scales, this picture of linear bias must be corrected due to halo exclusion and non-linear biasing effects \citep[see \egnospace][]{Baldauf2016, Desjacques2018, Ivanov2020, Pandey2020,baldauf21}. Even the simple linear bias model renders the amplitude of the galaxy clustering correlation function useless for inferring cosmological information. This degeneracy between galaxy bias and the variance of matter density fluctuations is broken when studying the full shape of the probability density function (PDF) of galaxy density fluctuations \citep{Uhlemann2018a, Friedrich2018, Repp2020}. However, analysing the full PDF shape comes with the additional complication that one also has to understand the scatter between galaxy density and matter density fluctuations (shot-noise or stochasticity, see \eg \citealt{Friedrich2018, Gruen2018} for a PDF context or \citealt{Hamaus2010, Desjacques2018} for stochasticity in 2-point statistics). Both for 2-point and PDF statistics recent analyses had to employ quite complex models of the stochastic relation between matter density and galaxy density (\eg \citealt{Friedrich2018,Gruen2018} using one parameter for galaxy bias and 2 parameters for density dependent shot-noise, \citealt{Uhlemann2018a} using 3 parameters to describe a function relating the cumulative distribution function of matter and galaxy density fluctuations and \citealt{Ivanov2020} using 3 parameters for galaxy bias and one shot-noise amplitude). 

In the PDF context, the bias of halos (or galaxies) \wrt the matter density field is typically incorporated through an Eulerian expansion of the conditional expectation value $\langle \delta_{\mathrm{halo}}|\delta_m\rangle$ (see \eg \citealt{Efstathiou1995, Manera2011, Clerkin2017, Friedrich2018, Gruen2018, Salvador2019, Repp2020}, with an exception found in \citealt{Uhlemann2018c}). This is somewhat unnatural, because standard methods to model the matter density PDF are typically built around the symmetric collapse of a leading order (saddle-point) configuration of the density field \citep[\egnospace][]{Bernardeau1994, Bernardeau1995, Valageas2002, Bernardeau2015, Uhlemann2016, Uhlemann2018b, Friedrich2020} which would seem to suggest a Lagrangian point-of-view.

We implement such a Lagrangian model in \secref{theory}, where we also give a general overview of PDF modelling and also review the non-Poissonian shot-noise model of \citet{Friedrich2018, Gruen2018} (hereafter F18 and G18). \Secref{simulations} presents details of the simulated data used in this study and in \secref{results} we assess the importance of different aspects of our theory, by comparing our model of the joint PDF $p(\delta_m, \delta_g)$ to the corresponding measured distribution of matter density and galaxy density fluctuations in those simulations. In particular, we are fitting both the Lagrangian and Eulerian bias models to measurements of $\langle \delta_{\mathrm{halo}}|\delta_m\rangle$ in simulated data at different redshifts, for different smoothing scales and using halos in different mass bins. We check whether the Lagrangian and Eulerian best-fitting parameters conform to consistency relations that should hold between them, and we compare them to corresponding values obtained from 2-point function measurements and from analytical predictions of bias as a function of halo mass. \Secref{shot-noise_details} investigates details concerning shot-noise of tracer density fields.  We discuss our results, summarize open questions and give an outlook on future work in \secrefnospace{conclusions}.

Throughout this paper, we consider the matter density and galaxy density fields averaged over cylindrical apertures (as opposed to \eg spherical ones). This makes our results more directly applicable to line-of-sight projections of the cosmic density fields, since PDF-related statistics of such projected fields are most efficiently expressed in terms of line-of-sight integrals of corresponding cylindrical quantities (\cf \citealt{Bernardeau2000, Friedrich2018, Barthelemy2019, Boyle2021}; this is the equivalent of the Limber approximation - \citealt{Limber1953} - for 2-point statistics). But our results can be easily transferred to the 3-dimensional case and to spherical apertures.

\section{Bias in the language of PDF cosmology}
\label{sec:theory}

\subsection{Galaxy bias from the joint cumulant generating function of matter and galaxy density}

In the following let $\delta_{m,R,L}(\bm{x}, z)$ and $\delta_{g,R,L}(\bm{x}, z)$ respectively be the matter and galaxy density contrast at redshift $z$ and location $\bm{x}$ when averaging over a cylindrical aperture of radius $R$ and length $L$ (the orientation of the cylinder does not play a role in the following due to statistical isotropy). In a statistically homogeneous and isotropic Universe, local moments of the form
\begin{equation}
    \langle \delta_{m,R,L}(\bm{x}, z)^k\ \delta_{g,R,L}(\bm{x}, z)^l \rangle
\end{equation}
do not depend on the spatial location $\bm{x}$ and we can define the joint moment generating function of matter and galaxy density contrast as
\begin{equation}
    \psi_{R,L}(\lambda_m, \lambda_g, z) \equiv \sum_{k,l\geq 0} \langle \delta_{m,R,L}(\bm{x}, z)^k\ \delta_{g,R,L}(\bm{x}, z)^l \rangle \frac{\lambda_m^k \lambda_g^l}{k!\ l!}\ .
\end{equation}
As evident from this definition, moments are obtained as derivatives of that function evaluated at $\lambda_m = 0 = \lambda_g$. For the rest of this sub-section, we will suppress any dependencies of our notation on $\bm{x}$, $z$, $R$ and $L$. From the moment generating function $\psi$ we define the cumulant generating function (CGF) as
\begin{align} 
    \varphi(\lambda_m, \lambda_g) \equiv&\ \log(\psi(\lambda_m, \lambda_g))\nonumber \\
    \equiv&\ \sum_{k,l\geq 1} \langle \delta_{m}^k\ \delta_{g}^l \rangle_c \frac{\lambda_m^k \lambda_g^l}{k!\ l!}\ ,
\end{align}
where the last line serves as a definition of the connected moments (or cumulants) $\langle \delta_{m}^k\ \delta_{g}^l \rangle_c$.

One quantity of interest for our study is the bias between galaxy density and matter density contrast as encoded by the conditional expectation value
\begin{equation}
    \langle \delta_{g} | \delta_{m} \rangle = \frac{1}{p(\delta_{m})} \int \dd \delta_{g}\ \delta_{g}\ p(\delta_{g} , \delta_{m})\ .
\end{equation}
Here $p(\delta_{m})$ is the probability density function (PDF) of matter density contrast $\delta_{m}$ and $p(\delta_{g} , \delta_{m})$ is the joint PDF of both $\delta_{g}$ and $\delta_{m}$ (at the same location and redshift and averaged over the same cylindrical aperture). This joint PDF is related to the CGF via an inverse Laplace transformation \citep[see \egnospace][]{Bernardeau2000, Valageas2002, Bernardeau2015, Friedrich2018}. Hence, the above expectation value can be computed as
\begin{align}
\label{eq:conditional_expectation_value}
    &\ \langle \delta_g | \delta_m \rangle \nonumber \\
    =&\ \frac{1}{p(\delta_m)} \int \frac{\dd\lambda_g\dd\lambda_m}{(2\pi)^2}\ e^{ - i\lambda_m\delta_m + \varphi(i\lambda_m, i\lambda_g)} \int \dd \delta_g\ \delta_g\ e^{-i\lambda_g\delta_g} \nonumber \\
    =&\ \frac{1}{p(\delta_m)} \int \frac{\dd\lambda_g\dd\lambda_m}{2\pi}\ e^{ - i\lambda_m\delta_m + \varphi(i\lambda_m, i\lambda_g)}\ i\frac{\dd \delta_{\mathrm{Dirac}}(\lambda_g)}{\dd \lambda_g} \nonumber \\
    =&\ \frac{\int \frac{\dd\lambda_m}{2\pi}\ e^{ - i\lambda_m\delta_m + \varphi(i\lambda_m)}\ \left.\partial_{l_g}\varphi(l_m, l_g)\right|_{l_m=i\lambda_m\ ,\ l_g=0}}{\int \frac{\dd\lambda_m}{2\pi}\ e^{ - i\lambda_m\delta_m + \varphi(i\lambda_m)}}\ ,
\end{align}
where $\varphi(\lambda_m)$ is the CGF of $\delta_m$ alone.

\subsection{The joint cumulant generating function from functional integration}
\label{sec:cylinder_CGF_with_bias}

To calculate $\langle \delta_g | \delta_m \rangle$ according to \eqnref{conditional_expectation_value} we need to know the joint CGF $\varphi_{R}(\lambda_g, \lambda_m)$, where we have re-introduced the dependence on the radius $R$ of our smoothing aperture, since we will occasionally vary $R$. The CGF can be calculated from the joint PDF as \citep{Bernardeau2015}
\begin{align}
\label{eq:CGF_as_expectation_value}
&\ e^{\varphi_{R}(\lambda_m, \lambda_g)} \nonumber \\
=&\ \langle e^{\lambda_g \delta_{g,R}+\lambda_m \delta_{m,R}} \rangle \nonumber \\
=&\ \int \dd \delta_{g,R}\ \dd \delta_{m,R}\ p(\delta_{g,R}, \delta_{m,R})\ e^{\lambda_g \delta_{g,R}+\lambda_m \delta_{m,R}}\ .
\end{align}
We want to stress again, that our smoothing apertures are cylindrical, \ie $R$ is the radius of these cylinders. The only reason for our use of cylindrical filtering is that we prepare for an analysis in line-of-sight projected data (Friedrich et al.\ in prep). And the CGF of a line-of-sight projected density field can be calculated in a Limber-type approximation \citep{Limber1953, Bernardeau2000, Friedrich2018, Barthelemy2019} from the CGF of the 3D density field in cylindrical apertures. But the following derivations apply in an almost identical manner to spherical filters as well.

Let us assume that both the galaxy density and matter density field are completely determined by the initial density field, or equivalently: today's linear density field which is related to the initial density field through linear growth. Then the expectation value in \eqnref{CGF_as_expectation_value} can also be expressed through a functional integral over all possible configurations of the linear density contrast \citep{Valageas2002}. This yields
\begin{align}
\label{eq:Valageas_functional_integral}
    e^{\varphi_{R}(\lambda_m, \lambda_g)} =\ & \int \mathcal{D}\delta_{\mathrm{lin}}\ e^{\lambda_g \delta_{g,R}[\delta_{\mathrm{lin}}]+\lambda_m \delta_{m,R}[\delta_{\mathrm{lin}}]}\ \mathcal{P}[\delta_{\mathrm{lin}}]\ ,
\end{align}
where $\delta_{g,R}[\cdot]$ and $\delta_{m,R}[\cdot]$ are now functionals and $\mathcal{P}[\cdot]$ is the probability density functional of the random field $\delta_{\mathrm{lin}}(\bm{x})$. For Gaussian initial conditions $\mathcal{P}[\cdot]$ is a Gaussian functional and determined completely by the linear power spectrum \citep{Valageas2002}. By re-expressing the probability density functional of $\delta_{\rm lin}$ in terms of its cumulant generating functional, \eqnref{Valageas_functional_integral} can be brought into a more general - and for our purposes more convenient - form. We thus follow \citet{Friedrich2020} who derived that
\begin{align}
\label{eq:Friedrich_functional_integral}
    e^{\varphi_{R}(\lambda_m, \lambda_g)} =\ & \frac{1}{\mathcal{N}} \int \mathcal{D}\delta_{\mathrm{lin}}\ \mathcal{D}J_{\mathrm{lin}}\ e^{-S_{\lambda_m, \lambda_g}[\delta_{\mathrm{lin}}, J_{\mathrm{lin}}]}\ ,
\end{align}
where $J_{\rm lin}$ is an auxiliary source associated with the initial conditions and the action $S_{\lambda_m, \lambda_g}[\delta_{\mathrm{lin}}, J_{\mathrm{lin}}]$ is defined as
\begin{align}
\label{eq:definition_of_action}
    &\ S_{\lambda_m, \lambda_g}[\delta_{\mathrm{lin}}, J_{\mathrm{lin}}]\nonumber \\
    \equiv&\  - \lambda_m \delta_{m,R}[\delta_{\mathrm{lin}}] - \lambda_g \delta_{g,R}[\delta_{\mathrm{lin}}] + iJ_{\mathrm{lin}} \cdot \delta_{\mathrm{lin}} - \Phi[iJ_{\mathrm{lin}}]\ .
\end{align}
Here $\Phi[\cdot]$ is now the cumulant generating functional of the random field $\delta_{\mathrm{lin}}$, and $\mathcal{N}$ is an irrelevant normalisation constant that drops in our final result \citep[\cfnospace][]{Friedrich2020}.

For detailed analyses of these and related functional integrals we \eg refer the reader to \citet{Valageas2002, Valageas2002V, Ivanov2019, Friedrich2020}. For the purpose of our study, we only state that the saddle point approximation to \eqnref{Friedrich_functional_integral} yields
\begin{equation}
\label{eq:saddle_point_approx_of_CGF}
\varphi_{R}(\lambda_m, \lambda_g) \approx -S_{\lambda_m, \lambda_g}[\delta_{\mathrm{lin}}^*, J_{\mathrm{lin}}^*]\ ,
\end{equation}
where $\delta_{\mathrm{lin}}^*$ and $J_{\mathrm{lin}}^*$ are the saddle point configurations of the fields $\delta_{\mathrm{lin}}(\bm{x})$ and $J_{\mathrm{lin}}(\bm{x})$ which minimise the action $S_{\lambda_m, \lambda_g}[\cdot , \cdot]$ and hence give the largest contribution to the functional integral. These saddle point configurations can be shown to exhibit the same symmetry as the aperture used to define the functionals $\delta_{g,R}[\cdot]$ and $\delta_{m,R}[\cdot]$ \citep{Valageas2002, Friedrich2018, Friedrich2020}. In the case of long cylindrical apertures ($L \gg R$) this means that $\delta_{\mathrm{lin}}^*$ and $J_{\mathrm{lin}}^*$ will be cylindrically symmetric functions. They can even be explicitly calculated \citep{Valageas2002, Friedrich2020}, which is however not needed for our purposes. What is more important is the fact that the functional $\delta_{m,R}[\cdot]$ can be easily determined in the cylindrically symmetric situation. If we are observing the density field at redshift $z$ then $\delta_{m,R}[\delta_{\mathrm{lin}}^*]$ is given by
\begin{equation}
\delta_{m,R}[\delta_{\mathrm{lin}}^*] = \mathcal{F}(\delta_{\mathrm{lin}, R_{\mathrm{lin}}}^*, z)\ .
\end{equation}
Here $R_{\mathrm{lin}}$ is the initial (Lagrangian) radius of all the matter that is enclosed within $R$ at redshift $z$, and $\delta_{\mathrm{lin}, R_{\mathrm{lin}}}^*$ is the average value of the saddle point configuration $\delta_{\mathrm{lin}}^*$ within this radius. Because of mass conservation $R_{\mathrm{lin}}$ is given by the (implicit) equation
\begin{equation}
    R_{\mathrm{lin}} = R \sqrt{1+\mathcal{F}(\delta_{\mathrm{lin}, R_{\mathrm{lin}}}^*, z)}\ ,
\end{equation}
and the function $\mathcal{F}(\delta_{\mathrm{lin}, R_{\mathrm{lin}}}^*, z)$ describes how a cylindrically symmetric perturbation evolves when today's linear density contrast within its initial radius is $\delta_{\mathrm{lin}, R_{\mathrm{lin}}}^*$. We detail the equations of motion needed to calculate $\mathcal{F}$ in \apprefnospace{cylindrical_collapse}.

So far we have reviewed existing results on calculating the cumulant generating function and extended the notation of \citealt{Friedrich2020} to the joint CGF of both galaxy density and matter density fluctuations as well as to cylindrical apertures instead of spherical ones. We will now see how the saddle point approximation of \eqnref{saddle_point_approx_of_CGF} allows for a practical implementation of a Lagrangian bias model within PDF theory.

\subsection{Lagrangian bias along the saddle point configuration}
\label{sec:Lagrangian_model}

\begin{figure}
  \includegraphics[width=0.47\textwidth]{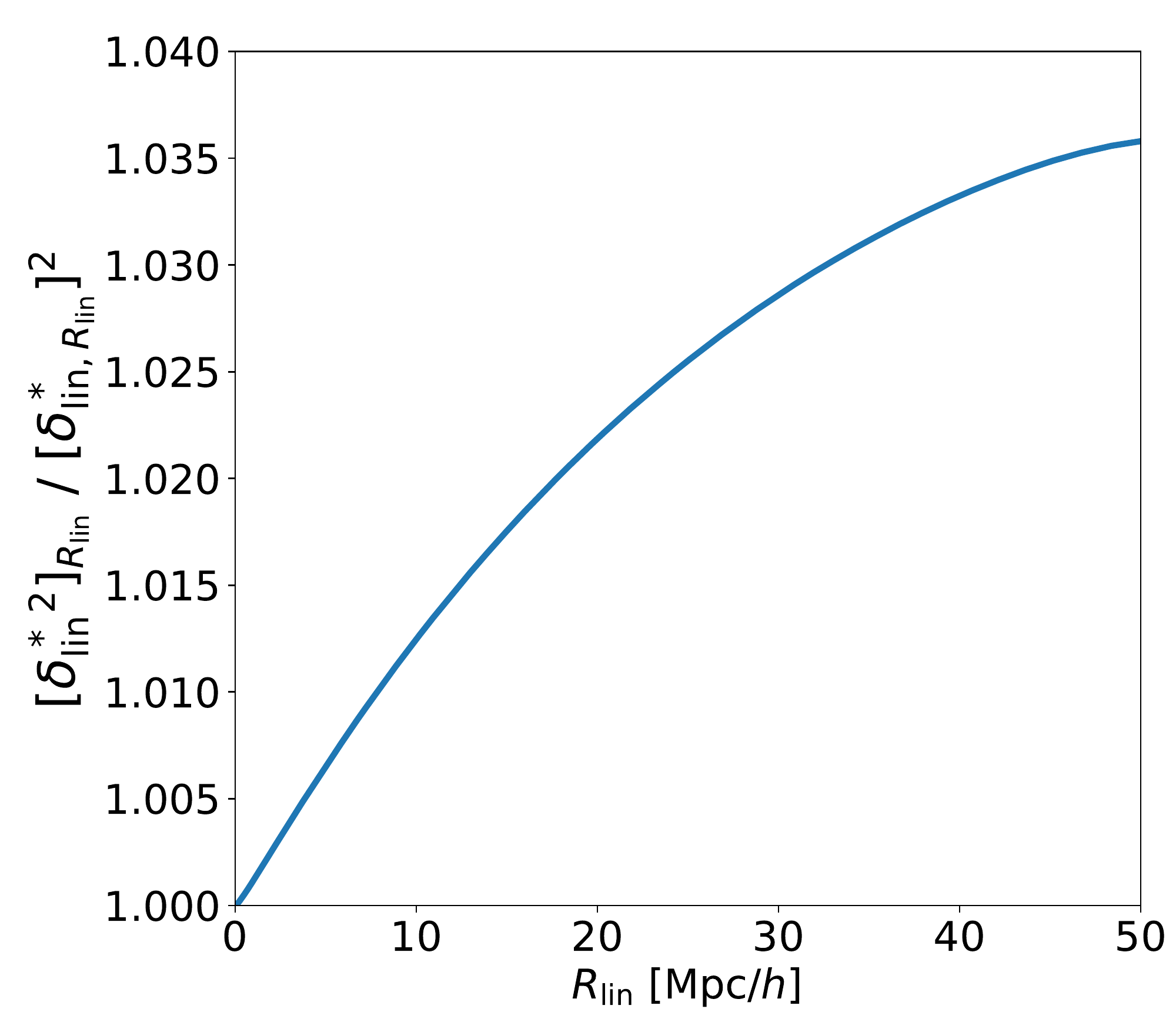}
   \caption{At the saddle point configuration which dominates the path integral of \eqnref{Friedrich_functional_integral} the operations of squaring and filtering the linear density contrast field commute approximately.}
  \label{fi:commutation}
\end{figure}

To implement a parametric model for halo bias, let us have a closer look at the functional $\delta_{g,R}[\delta_{\mathrm{lin}}]$. Since the saddle point configuration is cylindrically symmetric, we will only consider cylindrically symmetric configurations and effectively consider 2-dimensional density fields. If $\delta_g(\bm{r})$ is the (smooth, shot-noise free) galaxy density contrast at (the 2-dimensional) location $\bm{r}$, then $\delta_{g,R}$ is given by
\begin{equation}
    \delta_{g,R} = \frac{1}{\pi R^2} \underset{|\bm{r}|\leq R}{\int} \dd^2 r\ \delta_g(\bm{r})\ .
\end{equation}
Tracing back the cylindrically collapsing evolution of the saddle point, a mass element at location $\bm{r}$ will originate from some initial (Lagrangian) location $\bm{q}$. Following standard Lagrangian bias parametrizations \citep[see \egnospace][]{Lazeyras2016, Desjacques2018} we assume $\delta_g(\bm{r})$ can be expressed in terms of both the linear and non-linear matter density contrast field as
\begin{equation}
    1+\delta_g(\bm{r}) = (1+\delta_m(\bm{r}))\left(1+b_1^L\delta_{\mathrm{lin}}(\bm{q}) + \frac{b_2^L}{2}\delta_{\mathrm{lin}}(\bm{q})^2\right)\ ,
\end{equation}
where we have stopped the bias expansion at quadratic order in today's linear density contrast. The cylindrical average $1+\delta_{g,R}$ is then given by
\begin{align}
    &\ \frac{1}{\pi R^2} \underset{|\bm{r}|\leq R}{\int} \dd^2 r\ (1+\delta_m(\bm{r}))\left(1+b_1^L\delta_{\mathrm{lin}}(\bm{q}) + \frac{b_2^L}{2}\delta_{\mathrm{lin}}(\bm{q})^2\right)\nonumber \\
    =&\ \left(\frac{R_{\mathrm{lin}}}{R}\right)^2 \frac{1}{\pi R_{\mathrm{lin}}^2} \underset{|\bm{q}|\leq R_{\mathrm{lin}}}{\int} \dd^2 q\ \left(1+b_1^L\delta_{\mathrm{lin}}(\bm{q}) + \frac{b_2^L}{2}\delta_{\mathrm{lin}}(\bm{q})^2\right)\nonumber \\
    =&\ \left(\frac{R_{\mathrm{lin}}}{R}\right)^2 \left( 1 + b_1^L \delta_{\mathrm{lin}, R_{\mathrm{lin}}} + \frac{b_2^L}{2} [\delta_{\mathrm{lin}}^2]_{R_{\mathrm{lin}}} \right)\ .
\end{align}
Here $R_{\mathrm{lin}}$ is again the initial, Lagrangian (or \emph{linear}) radius of the cylindrical perturbation now enclosed within $R$, $\delta_{\mathrm{lin}, R_{\mathrm{lin}}}$ is the average of today's linear density contrast within $R_{\mathrm{lin}}$ and $[\delta_{\mathrm{lin}}^2]_{R_{\mathrm{lin}}}$ is the average of the squared linear density contrast within $R_{\mathrm{lin}}$. Since we are considering cylindrically collapsing perturbations, the Lagrangian radius $R_{\mathrm{lin}}$ is related to $R$ through
\begin{equation}
    R_{\mathrm{lin}} = R \sqrt{1 + \delta_{m,R}}\ .
\end{equation}
Hence, $\delta_{g,R}$ within our quadratic Lagrangian bias model is given by
\begin{equation}
    1+\delta_{g,R} = (1+\delta_{m,R}) \left( 1 + b_1^L \delta_{\mathrm{lin}, R_{\mathrm{lin}}} + \frac{b_2^L}{2} [\delta_{\mathrm{lin}}^2]_{R_{\mathrm{lin}}} \right)\ .
\end{equation}
In \figref{commutation}, which is based on calculations presented in our \apprefnospace{commutation}, we show that for the saddle point configuration $\delta_{\mathrm{lin}}^*$ the operations of squaring and cylindrically averaging approximately commute, \ie
\begin{equation}
    [{\delta_{\mathrm{lin}}^*}^2]_{R_{\mathrm{lin}}} \approx (\delta_{\mathrm{lin}, R_{\mathrm{lin}}}^*)^2\ .
\end{equation}
This allows us to express $\delta_{g,R}[\cdot]$ along the saddle point as
\begin{align}
\label{eq:bias_at_saddle_point}
    &\ 1+\delta_{g,R}[\delta_{\mathrm{lin}}^*, z] \approx \nonumber\\
    &\ (1+\delta_{m,R}[\delta_{\mathrm{lin}}^*, z])\left(1 + b_1^L\ \delta_{\mathrm{lin}, R_{\mathrm{lin}}}^* + \frac{b_2^L}{2}(\delta_{\mathrm{lin}, R_{\mathrm{lin}}}^*)^2\right)\ .
\end{align}
We now have all the ingredients to formulate our main technical result. In complete analogy to the derivations of \citealt{Friedrich2020} (but for cylindrical apertures and using the modified action of \eqnrefnospace{definition_of_action}) the task of determining the saddle point value of the action, $S_{\lambda_m, \lambda_g}[\delta_{\mathrm{lin}}^*, J_{\mathrm{lin}}^*]$, is equivalent to minimising the 2-dimensional function 
\begin{align}
    & s_{\lambda_m, \lambda_g}(\delta, j) = \nonumber \\
    &  - \lambda_g(1+\mathcal{F}(\delta, z)) \left(b_1^L \delta + \frac{b_2^L}{2} \delta^2\right) - (\lambda_m + \lambda_g)\ \mathcal{F}(\delta, z) \nonumber \\
    & + j \delta - \varphi_{{\mathrm{lin}}, R(1+\mathcal{F}(\delta, z))^{1/2}}(j)\ .
\end{align}
Here $\varphi_{{\mathrm{lin}}, R}$ is the CGF of the linear density contrast (which is a quadratic function for Gaussian initial conditions) and $\delta$ and $j$ should be understood as scalar variables. Minimising $s_{\lambda_m, \lambda_g}(\delta, j)$ \wrt these variables yields an approximation of the joint CGF of matter density and galaxy density fluctuations via \eqnrefnospace{saddle_point_approx_of_CGF}. This is the main result of our paper. Our formalism based on functional integration would be equivalent to a derivation within large deviation theory \citep[LDT, see][who introduced LDT for the matter density PDF]{BernardeauReimberg2016}, so we will refer to our calculation as the LDT model.

In practice we enhance the accuracy of this approximation with a linear-to-nonlinear variance re-scaling of the cumulant generating function that leaves the reduced cumulants $S_n \equiv \langle \delta_m^n\rangle_c/\langle \delta_m^2\rangle_c^{n-1}$ unchanged (see \eg Section IV.A.2 of \citealt{Friedrich2018}). This however doesn't affect first derivatives of the CGF and has hence little impact on our calculation of $\langle \delta_g|\delta_m\rangle$ via \eqnrefnospace{conditional_expectation_value}. Numerical implementation of the minimisation of $s_{\lambda_m, \lambda_g}(\delta, j)$ can be achieved in a manner similar to the one detailed step-by-step in section 4.6 of \citet{Friedrich2020}. Equipped with the above approximation for the cumulant generating function we are now in a position to evaluate \eqnref{conditional_expectation_value} and hence calculate the expectation value $\langle \delta_g | \delta_m \rangle$. In the following we will compare this Lagrangian bias model to an Eulerian model, which we directly define as a Taylor expansion of $\langle \delta_{g}|\delta_m\rangle$, \ie
\begin{equation}
\label{eq:Eulerian_expec}
    \langle \delta_{g}|\delta_m\rangle = b_1^\mathrm{E}\ \delta_m + \frac{b_2^\mathrm{E}}{2} \left(\delta_m^2 - \langle \delta_m^2\rangle\right)\ .
\end{equation}
This parametrization ignores tidal bias terms that can also contribute at second order in $\delta_m$ \citep[\egnospace][]{Baldauf2012, Desjacques2018}. Since we are averaging over cylindrical apertures we expect these contributions to partially average out for the filtered density contrast \citep[\cf figure 3 of][]{Baldauf2012} but our best-fitting values for $b_2^{\mathrm{E}}$ may absorb residual tidal contributions and hence may be slightly biased. We do not investigate this here. Subtracting the constant term $b_2^\mathrm{E}/2\cdot\langle \delta_m^2\rangle$ in \eqnref{Eulerian_expec} ensures that $\langle \delta_{g}\rangle = 0$. Note that this is not necessary in our Lagrangian model because of the Lagrangian-to-Eulerian mapping that is built into our path integral formulation.

\subsection{Non-Poissonian shot-noise}
\label{sec:non_Poissonian_shot_noise_model}

The joint PDF of $\delta_m$ and $\delta_{g}$ can be expressed as
\begin{equation}
    p(\delta_m,\delta_{g}) = p(\delta_m)\ p(\delta_{g}|\delta_m)\ .
\end{equation}
The matter density PDF $p(\delta_m)$ appearing on the right hand side of this equation can be computed as the inverse Laplace transform of the cumulant generating function of $\delta_m$ (\cf the denominator in the last line of \eqnref{conditional_expectation_value} as well as \citealt{Valageas2002, Bernardeau2015, Friedrich2018, Friedrich2020} for practical implementations of that transform). The second factor of the above equation, $p(\delta_{g}|\delta_m)$, is the conditional PDF of tracer density fluctuations given a fixed value of $\delta_m$. In the previous subsections we have focused on computing the expectation value of that distribution, $\langle\delta_{g}|\delta_m\rangle$.

To model the full distribution $p(\delta_{g}|\delta_m)$ we have to consider stochasticity (\resp shot-noise) around the expectation value $\langle\delta_{g}|\delta_m\rangle$. This noise is often assumed to be Poissonian \citep[see \eg][]{Efstathiou1995, Clerkin2017, Salvador2019, Repp2020}. However, the results of F18 and G18 indicate that for certain types of tracers \citep[in their case luminous red galaxies; \cfnospace][]{Rozo2016} this assumption can be in inaccurate \citep[see also][for non-Poissonian shot-noise in different contexts]{Hamaus2011, Dvornik2018}. To account for deviations from Poisson noise, F18 and G18 have modelled the distribution of a discrete random variable $N$ with expectation value $\bar N$ as
\begin{equation}
    P_\alpha(N) = \mathcal{N} \exp\left\lbrace \frac{N}{\alpha}\ln\left[\frac{\bar N}{\alpha}\right] - \ln \Gamma \left[\frac{N}{\alpha}+1\right] - \frac{\bar N}{\alpha}\right\rbrace\ .
\end{equation}
Here $\alpha$ parametrizes deviations from Poisson noise (with $\alpha=1$ leading to a Poisson distribution), $\Gamma$ is the gamma-function and $\mathcal{N}$ is a normalisation factor. That normalisation is to a good approximation given by $1/\alpha$, though we do not rely on this here.

The above ansatz for $P_\alpha(N)$ can be used to model the distribution of tracer counts $N_{g}$ in an aperture filled with a matter density contrast $\delta_m$ if we perform the identifications
\begin{align}
    N \rightarrow &\ N_{g} \nonumber\\
    \bar N \rightarrow &\ \bar N_{g}\ (1+ \langle \delta_{g} | \delta_m\rangle) \nonumber\\
    P_\alpha(N) \rightarrow &\ P_\alpha(N_{g}|\delta_m)\ , \nonumber
\end{align}
where $\bar N_{g}$ is the mean tracer count across all apertures in a given survey volume. F18 and G18 then allow $\alpha$ to be a function of $\delta_m$ as well, hence making deviations from Poisson noise a function of the underlying matter density. They found that a linear ansatz,
\begin{equation}
    \alpha(\delta_m) = \alpha_0 + \alpha_1 \delta_m
\end{equation}
describes the redMaGiC galaxy sample of the Buzzard N-body simulations \citep{DeRose2017} well. We will test this linearity assumption here for a different set of simulations and different tracer samples of the large-scale structure.

\begin{figure*}
  \includegraphics[width=0.98\textwidth]{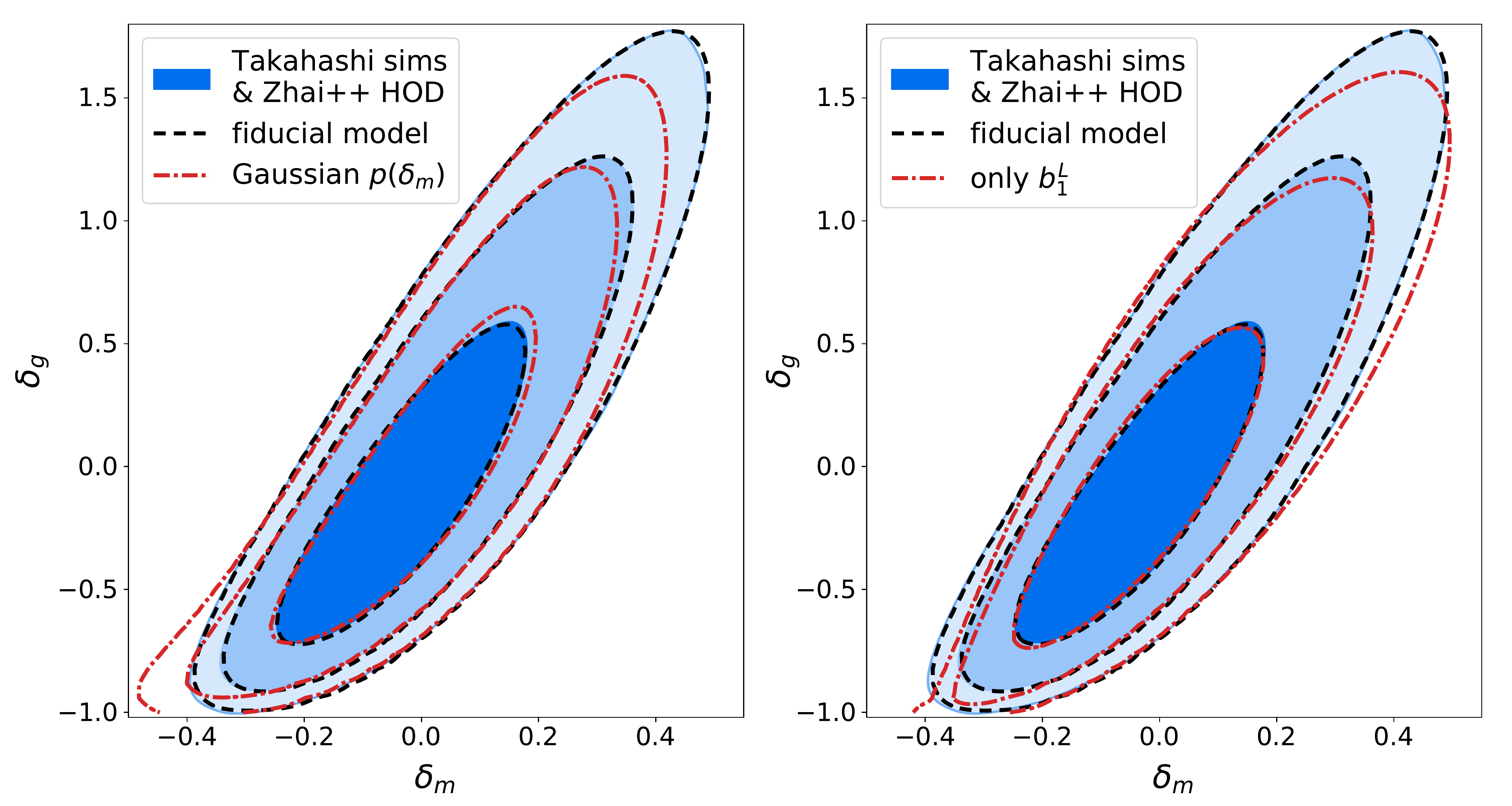}
  \caption{Comparing different models of the joint PDF $p(\delta_g, \delta_m)$ of galaxy and matter density fluctuations in cylindrical apertures of length $L=150$ Mpc$/h$ and radius $R=20$ Mpc$/h$ at redshift $z\approx 0.75$ to the distribution measured in simulated data. In both panels the blue contours represent the PDF measured in T17 mock data and using the mock galaxy catalog described in \secrefnospace{HOD}. The black, dashed contours represents our fiducial model, which consists of three parts: a large deviation theory (LDT) model for the matter density PDF $p(\delta_m)$, a second order Lagrangian bias expansion within LDT and a shot-noise model that allows for deviations from Poisson shot-noise (\cf \secref{non_Poissonian_shot_noise_model} for the shot-noise model, and \secref{shot-noise_details} for a detailed analysis of shot-noise in our simulations). The red, dash-dotted contours in the left panel show what happens to the joint PDF model, if one assumes that $p(\delta_m)$ is Gaussian. The red, dash-dotted contours in the right panel show a model that only fits a linear Lagrangian bias expansion.}
  \label{fi:joint_PDF}
\end{figure*}

\section{Simulated data}
\label{sec:simulations}

The following section presents details of the different simulated data sets we use to test the theoretical ansatzes of \secrefnospace{theory}.

\subsection{T17 N-body simulations}
\label{sec:T17_sims_description}

We use publicly available data from cosmological simulations run by \cite{Takahashi2017}\footnote{The data products of the simulation are available at \url{http://cosmo.phys.hirosaki-u.ac.jp/takahasi/allsky_raytracing/}.}. In the following we refer to these as the T17 simulations. The simulations were generated primarily for the gravitational lensing studies for the Hyper Suprime Cam Survey. In this paper, we use the full-sky light-cone halo catalogs and matter density contrast shells of the simulation suite.

These data sets were obtained from a cold dark matter (CDM) only cosmological N-body simulation in periodic cubic boxes. The simulations consist of 14 boxes of increasing side lengths $L, 2L, 3L, ..., 14L$ (with $L=450 \; \mathrm{Mpc/h}$), nested around a common vertex (see Figure 1 of \citealt{Takahashi2017}). Each box contains $2048^3$ particles (smaller boxes hence have better spatial and mass resolution) and their initial conditions were set with second-order Lagrangian perturbation theory \citep{Crocce2006} with an initial power spectrum computed for a flat $\Lambda$CDM cosmology with the following parameters : $\Omega_{cdm} = 0.233,\; \Omega_{b} = 0.046,\; \Omega_{m} = \Omega_{cdm} + \Omega_{b} = 0.279,\; \Omega_{\Lambda} = 0.721,\; h = 0.7,\; \sigma_8 = 0.82\; \mathrm{and}\; n_s = 0.97$. The particles in each box were then made to evolve from the initial conditions using the the N-body gravity solver \verb GADGET2  \citep{Springel2001, Springel2005}. Dark matter halos and sub-halos in each simulation box were identified using the six-dimensional phase-space friends-of-friends algorithm \verb ROCKSTAR \citep{Behroozi2013}. These \verb ROCKSTAR  halo catalogs and the evolved particle distribution of the different nested boxes are combined in layers of shells, each $150 \; \mathrm{Mpc/h}$ thick, to obtain full-sky light cone halo catalogs and matter density contrast inside the shells, respectively. The simulation boxes were also ray traced using the multiple-lens plane ray-tracing algorithm \verb GRAYTRIX  \citep{Hamana2015, Shirasaki2015} to obtain weak lensing convergence/shear maps for several source redshifts. Multiple simulations were run to produce 108 realizations (with labels r000 to r107 ) for each of these data products (see \cite{Takahashi2017} for more details). The authors report that the average matter power spectra from their several realizations of the simulations agreed with the theoretical revised \verb Halofit  power spectrum \citep{Smith2003, Takahashi2012} to within 5 (10) per cent for $k < 5 (6) \; \mathrm{h/Mpc}$  at $z < 1$.

In this paper, for studying the bias as a function of halo properties we use the matter density contrast and the identified halos in three $150 \; \mathrm{Mpc/h}$ thick shells centred at $z = 0.476, 0.751, 0.990$ of realization r000 of the simulation suite. The all-sky halo catalogs come with a variety of halo properties such as halo mass, positions etc. of which we make use of the halo positions (right ascension, declination and redshift), halo mass $M_{200b}$ (i.e. the mass contained in a radius within which the overdensity equals 200 times the background density), the virial radius of the halo $R_{vir}$ and the scale radius $R_s$, obtained by fitting an NFW profile \citep{NFW} to a given halo. The concentration parameter of the halo can then be calculated as $c \equiv R_{vir} / R_s$. Technically, our halo catalogs do contain sub-halos. But the sub-halo fraction is negligible ($<0.1\%$ of the total halo population for the shell at $z = 0.476$ and even smaller for the other shells) such that for all practical purposes all halos can be considered to be parent halos.

\subsection{Populating galaxies within T17 halos using an HOD approach}
\label{sec:HOD}

The T17 simulation suite does not come with galaxy catalogs. We would however like to validate our methods for typical luminous red galaxies (LRGs) similar to those observed by eBOSS \citep[\egnospace][$z\approx 0.7$]{Zhai2017, Ross2020}. We hence create our own full-sky mock galaxy catalog by populating the T17 halo catalog at $z=0.75$ using an empirical Halo Occupation Distribution (HOD) method \citep{Berlind_2002} based on the widely used halo model of large-scale structure (see \cite{Cooray2002} for a review). Briefly, an HOD describes a probability distribution $P(N_g|M_h)$, \ie the probability that a given halo of mass $M_h$ hosts $N_g$ galaxies of a specific type (e.g. eBOSS LRG like galaxies). We assume that the HOD does not depend on environment or formation history of the halos (also known as assembly bias). We follow the work of \cite{Zhai2017} who empirically studied the clustering of more than 97000 LRGs in the eBOSS survey within $z=0.6-0.9$ (which contains the redshift range of the shell centred at $z=0.75$) using a 5-parameter HOD (we refer to this as the Zhai HOD). \cite{Zhai2017} parametrize their HOD by separating the contribution of a central galaxy from that of the satellite galaxies in a given halo of mass $M_h$. They characterised these contributions using the following functional forms for the mean values of the central and satellite galaxies:
\begin{equation}
\label{eq:HOD_mean_N_central}
    \langle N_{cen} \vert M_h \rangle = \frac{1}{2} \Big[ 1 + \mathrm{erf} \left( \frac{\log M_h - \log M_{min}}{\sigma_{\log M_h}} \right)\Big] \ ,
\end{equation}
\begin{equation}
\label{eq:HOD_mean_N_satellite}
    \langle N_{sat} \vert M_h \rangle = \left( \frac{M_h}{M_{sat}}  \right)^{\gamma} \exp \left( -\frac{M_{cut}}{M_h} \right) \langle N_{cen} \vert M_h \rangle \ .
\end{equation}
The first of the above equations describes a smooth transition between having either 0 or 1 central galaxy with $M_{min}$ being the mass at which half the halos (in a given sample) host a central galaxy and $\sigma_{\log M_h}$ gives the scatter of the halo mass $M_h$ at a fixed galaxy luminosity. The second equation gives the mean occupancy of satellite galaxies within the halo and is further parametrized by $\gamma$ - a power-law index for the mass dependence of the number of satellites, $M_{sat}$ - threshold mass for halos to contain one satellite, and $M_{cut}$ which allows for a halo-mass dependent cutoff. Together, the mean number of galaxies hosted within a halo of mass $M_h$ is given by:
\begin{equation}
\label{eq:HOD_mean_N_galaxies}
    \langle N_g \vert M_h \rangle = \langle N_{cen} \vert M_h \rangle +  \langle N_{sat} \vert M_h \rangle \ .
\end{equation}
\cite{Zhai2017} provide their best-fitting values for the 5 parameters by fitting analytical correlation functions\footnote{the one-halo and two-halo correlation functions, see e.g. Appendix A of \cite{Coupon2012}.} written in terms of their HOD to the observed galaxy clustering 2-point correlation functions of the eBOSS LRGs sample. We report their best-fitting values here (see Table 2 of \citealt{Zhai2017}): $\log M_{min} = 13.67,\;  \log M_{sat} = 14.93,\; \gamma = 0.43,\; \log M_{cut} = 11.62,\; \sigma_{\log M_h} = 0.81$, where it is assumed that all the masses are expressed in units of $M_{\odot}/h$. In order to obtain these values \citet{Zhai2017} have adopted $M_{200b}$ as their halo mass definition and we do so as well throughout our paper.

In order to create our mock galaxy catalog from the T17 simulation, we use the \verb|halotools| software \citep{Hearin_2017} to first combine the T17 r000 halo shells which span the redshift range $z=0.6-0.9$ to obtain a halo catalog.
Using $M_{200b}$ as the mass proxy for the halo mass $M_h$, we use the Zhai HOD that we have described above along with their best-fitting parameters to populate each halo in the catalog with galaxies. Note however, that we restrict ourselves to halos with masses $M_{200b} > 7.4\cdot10^{12} \; M_{\odot}/h$ for the generation of our galaxy catalog. This is to ensure that we have a similar number density of mock galaxies (per arcmin$^2$) as reported by \cite{Zhai2017} in their Table 1 for the total BOSS+eBOSS LRG sample.\footnote{A more accurate approach would be to re-fit our HOD parameters by matching a sufficiently constraining set of statistics of our mock galaxies to a target observed galaxy sample. This is however beyond the scope of this work.} To this end, for a given halo we perform a Bernoulli draw with expectation given by equation \eqref{eq:HOD_mean_N_central} to get $N_{cen}$ and a Poisson random draw with expectation given by equation \eqref{eq:HOD_mean_N_satellite} to obtain $N_{sat}$. The halo is then assigned to have a count of $N_{cen} + N_{sat}$ galaxies, where the central galaxy is placed at the same location as that of the parent halo's coordinates whereas a given satellite galaxy is placed at a distance $r$ $\textrm{Mpc}$ from the centre of the halo where $r$ is a random realization\cprotect\footnote{Precisely, we use the \verb|mc_generate_nfw_radial_positions| method from \verb|halotools| to draw a satellite galaxy's radial location $r$ inside a given halo of mass $M_{200b}$, concentration parameter $c$ and redshift $z$.} of a point drawn from an NFW profile. Besides, the radial distance from the centre of the given halo, each satellite galaxy is assigned a uniformly distributed random angular direction on the sphere of radius $r$, from the centre of the halo. In this way, we create a mock full-sky eBOSS LRG like galaxy catalog which we use for our analysis.

\subsection{Quijote N-body simulations}
\label{sec:Quijote}

The Quijote suite of N-body simulations \citep{Navarro2019} have been developed for quantifying the cosmological information content of large scale structure observables. The suite consists of 43100 simulations evaluated for more than 7000 cosmological models, varying the standard $\Lambda$CDM parameters, $M_\nu$ and $w$. For our study we made use of the high-resolution runs of Quijote, which follow the evolution of $1024^3$ particles over a co-moving volume of 1 $({\rm Gpc}/h)^3$ starting from $z=127$ for a fixed fiducial cosmology. Snapshots and halo catalogs (generated using a friends-of-friends algorithm) are publicly available for redshifts $z=0,0.5,1,2,3$. Matter density PDFs are already included with the associated data products, and we extracted the joint tracer-matter PDFs. We refer the reader to \cite{Navarro2019} for further details. 

The Molino suite of mock galaxy catalogues has been 
created from the Quijote $N$-body simulations in order to 
extend cosmological forecasts to galaxy observables. 
The suite contains 75,000 mock galaxy catalogs that are 
constructed by applying the \cite{Zhai2017} HOD model ( Section~\ref{sec:HOD}) to the Quijote halo catalogs. 
The galaxy catalogues are available at multiple cosmologies necessary for Fisher matrix forecasts (though here we only use catalogs at the Quijote fiducial cosmology of $(\Omega_m, \Omega_b, \sigma_8, n_s, h) = (0.3175, 0.049, 0.834, 0.9624, 0.6711)$).

\section{Comparison of theory and simulated data}
\label{sec:results}

We now compare the theoretical ansatzes developed in \secref{theory} to the simulated data described in \secrefnospace{simulations}. We start in \Secref{T17_and_Zhai} by looking at the joint PDF of matter density and our T17 synthetic galaxy sample. In \secref{bias_details} we then investigate the performance of our bias models as a function of mass, scale and redshift. And in \secref{shot-noise_details} we have a more detailed look at the shot-noise of different kinds of tracer samples.

\subsection{The joint PDF of matter and galaxy density}
\label{sec:T17_and_Zhai}

In \figref{joint_PDF} we compare different models for the joint distribution of galaxy density and matter density fluctuations to a corresponding measurement of that distribution in the T17 simulations (\cf \secrefnospace{simulations}). The total matter density contrast of T17 is available in concentric shells of thickness $150$ Mpc/h. For \figref{joint_PDF} we choose the shell centred around $z\approx 0.75$, which is \eg similar to the average redshift of galaxy samples recently used in analyses of eBOSS \citep{Zhai2017, Bautista2020,Gil-Marin2020, deMattia2020, Tamone2020}. The redshifts of the eBOSS LRG and ELG (emission line galaxy) samples span ranges that are significantly wider than $150$ Mpc$/h$. Hence, the Limber-type approximation that one would employ when studying the line-of-sight projected PDF of these samples will not significantly deteriorate the accuracy we find here for the T17 shell width. To generate our mock galaxy sample we populates T17 halos with the HOD described by \citet[][\cf our \secrefnospace{HOD}]{Zhai2017}. To both the matter density and galaxy density map we then apply a circular top-hat filter with radius $R = 20$ Mpc$/h$ perpendicular to the line-of-sight, \ie we are averaging both fields in approximately cylindrical apertures of length $L=150$ Mpc$/h$ and radius $R=20$ Mpc$/h$.

The blue contours in the two panels of \figref{joint_PDF} represent $1\sigma$, $2\sigma$ and $3\sigma$ quantiles of the joint distribution $p(\delta_g, \delta_m)$ in our T17 + \citeauthor{Zhai2017} mock data. The black contours represent the same quantiles for the theoretical model of $p(\delta_g, \delta_m)$ presented in \secrefnospace{theory}. To obtain the Lagrangian bias parameters of that model, we have fit our theoretical prediction of $\langle \delta_g | \delta_m \rangle$ to measurements of that conditional expectation value in the simulated density fields. We performed these measurements in 25 equidistant bins of $\delta_m$ within a range that cuts $2\%$ of the probability from each tail of the PDF $p(\delta_m)$.

The red contours in the left panel of \figref{joint_PDF} show the theoretical distribution $p(\delta_g, \delta_m)$ that one would obtain when assuming that $p(\delta_m)$ is a Gaussian PDF (and hence solely determined by its variance). Clearly, such a description is not sufficient for the matter density field at the smoothing scales and redshift considered here. The red contours in the right panel of \figref{joint_PDF} show the distribution $p(\delta_g, \delta_m)$ that would be predicted when fitting only a linear Lagrangian bias model. Clearly, such a model does not sufficiently capture the curvature of $\langle \delta_g | \delta_m \rangle$ \wrt $\delta_m$.

\begin{figure}
  \includegraphics[width=0.48\textwidth]{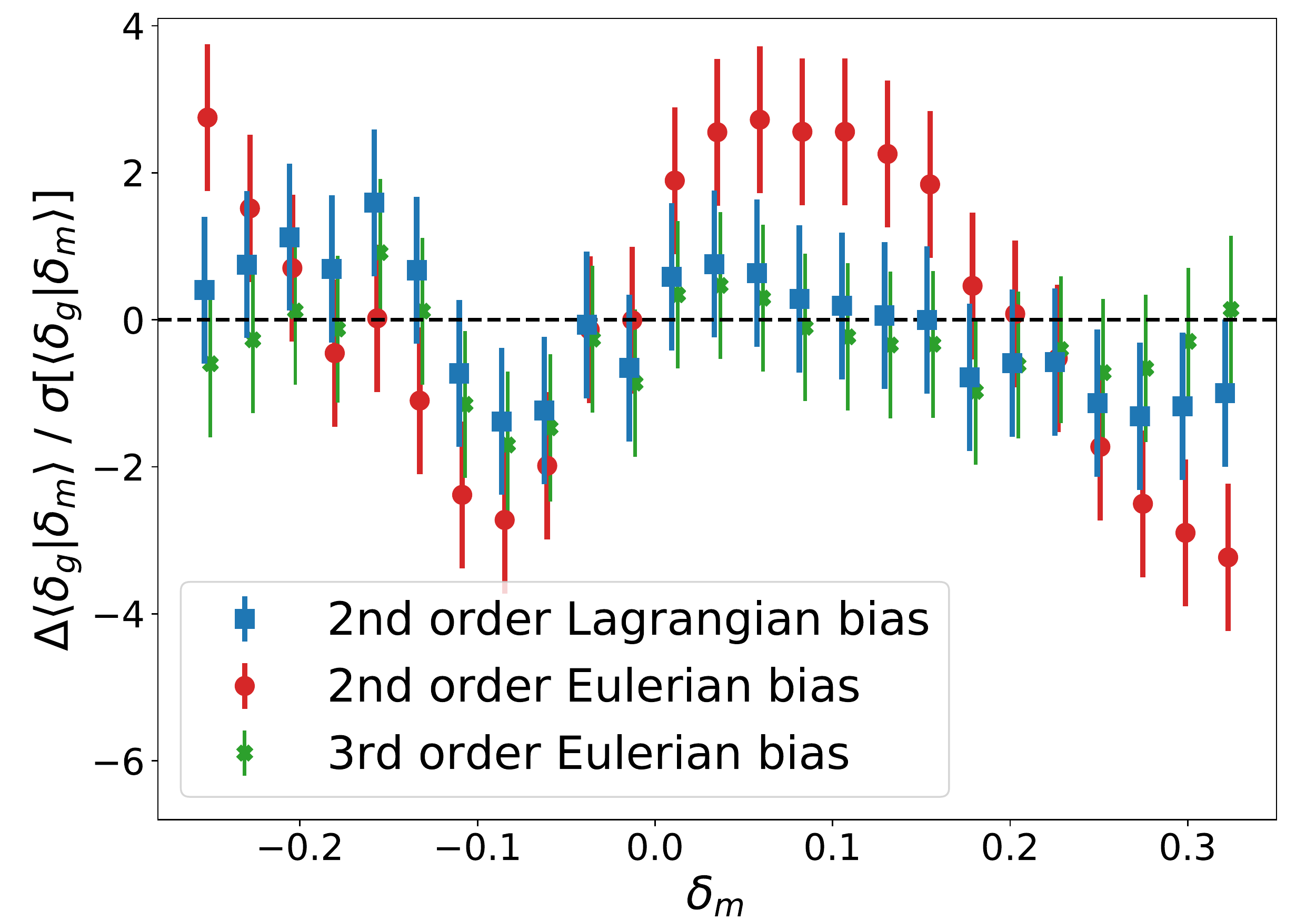}
  \caption{Residuals of $\langle \delta_g | \delta_m \rangle$ measured in simulated data (T17 halos populated with the HOD description of \citealt{Zhai2017}) \wrt our best-fitting, quadratic Lagrangian (blue squares) and Eulerian (red circles) models. The figure uses the same scales and redshift as \figref{joint_PDF} and the residuals have been normalised by an estimate of the standard deviation of our measurements of $\langle \delta_g | \delta_m \rangle$. We also show that a 3rd order Eulerian model (green crosses) performs similar to the second order Lagrangian one.}
  \label{fi:L_vs_E_Zhai}
\end{figure}

The difference between our best-fitting Lagrangian and Eulerian models for $p(\delta_g, \delta_m)$ is significantly smaller than the differences displayed in \figrefnospace{joint_PDF}. Hence we don't visualise them on the level of the full PDF, but for the conditional expectation values $\langle \delta_g | \delta_m \rangle$. \Figref{L_vs_E_Zhai} shows the residuals of $\langle \delta_g | \delta_m \rangle$ measured in our simulated data \wrt our best-fitting, quadratic Lagrangian and Eulerian model (blue squares and red circles; within the range used to fit both models which cuts $2\%$ of probability from the tails of $p(\delta_m)$). We normalise these residuals by the $1\sigma$ standard deviations estimated with a jackknife scheme (\cf \secref{bias_details} for more details). The Lagrangian model manages to achieve a significantly better fit to our simulated data than the Eulerian one. The figure also shows the residuals of a best-fitting cubic Eulerian model which adds a term $b_3^E/6 \cdot (\delta_m^3 - \langle \delta_m^3 \rangle)$ to \eqnrefnospace{Eulerian_expec}. This model performs very similar to the second order Lagrangian fit. Note however that the errorbars of \figref{L_vs_E_Zhai} represent all-sky data, \ie they might overestimate the accuracy required for realistic analyses and the second order Eulerian model may still perform well enough for those. Also, we find in \secref{bias_details} that this comparison is somewhat mass dependent: the Lagrangian model tends to perform better for intermediate mass halos, while the Eulerian one achieves better fits of $\langle \delta_g | \delta_m \rangle$ for very massive tracers of the density field.

\subsection{Halo bias as a function of mass and consistency among bias measures}
\label{sec:bias_details}

\begin{figure*}
  \includegraphics[width=0.98\textwidth]{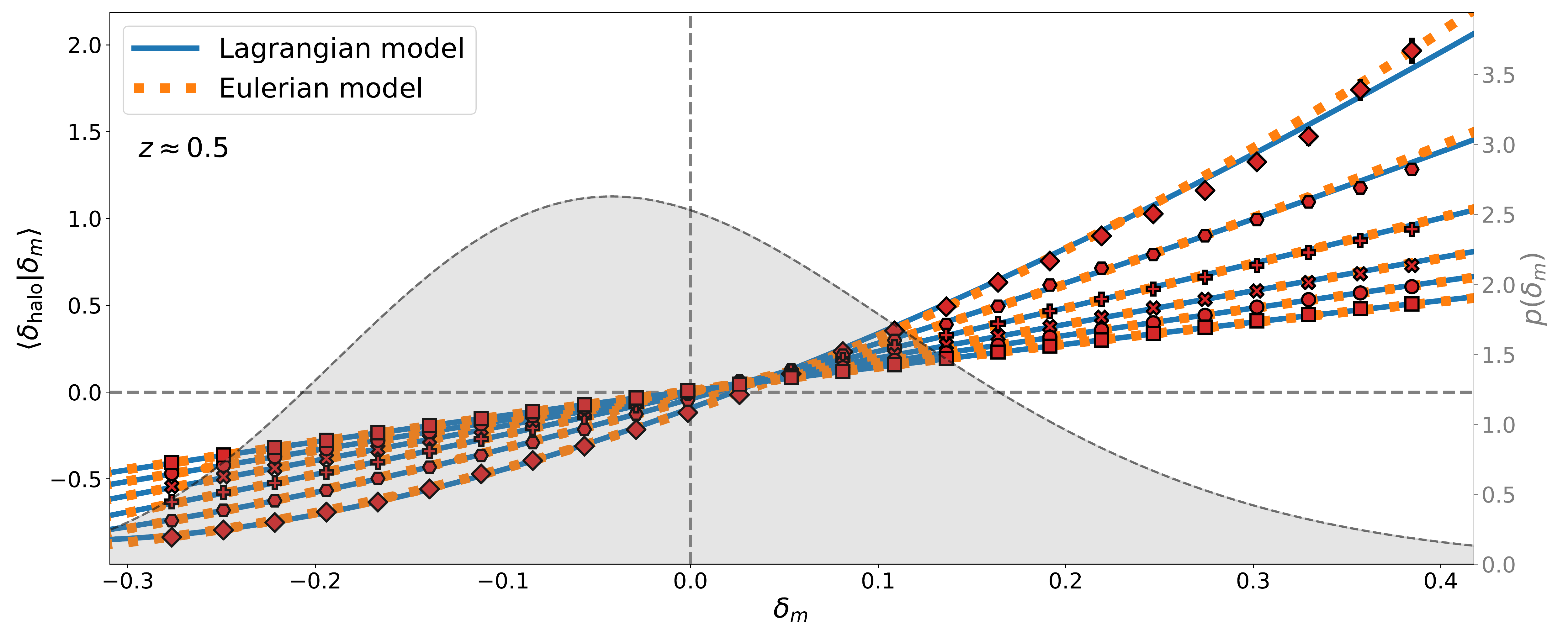}
  
  \includegraphics[width=0.98\textwidth]{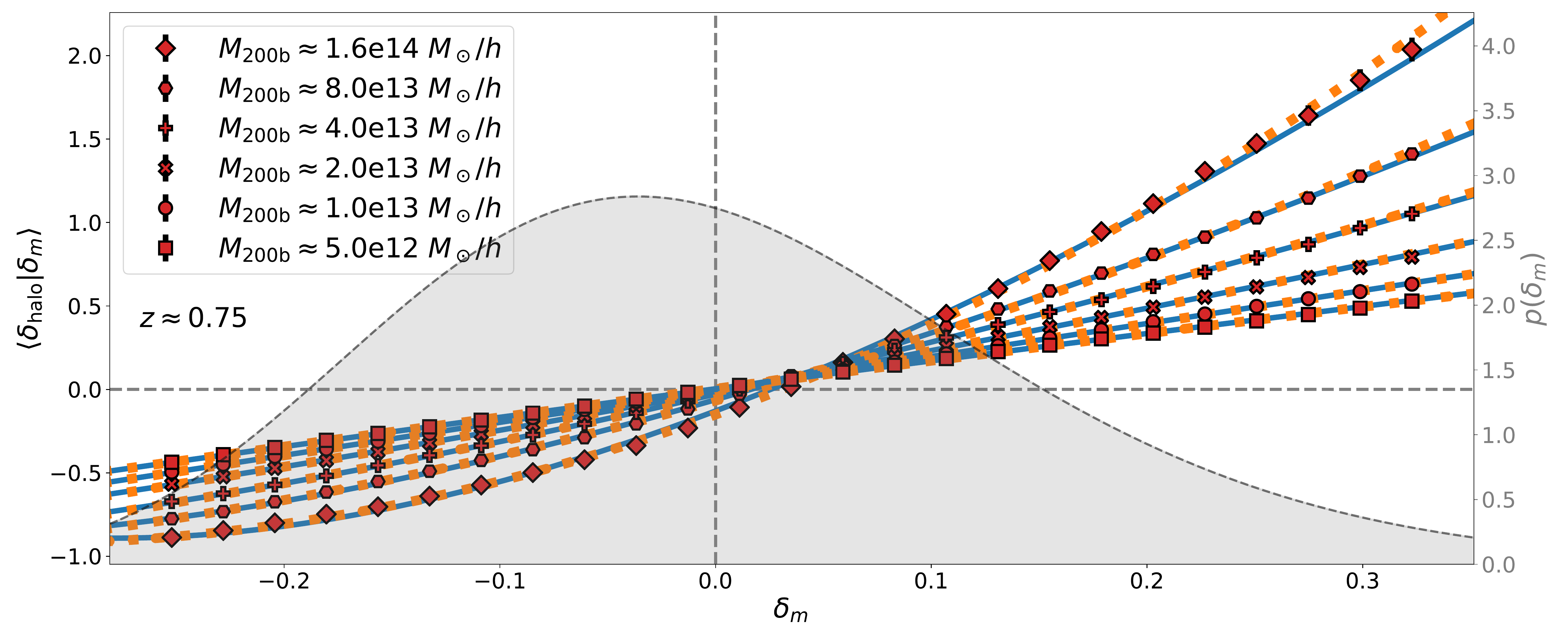}
  
  \includegraphics[width=0.98\textwidth]{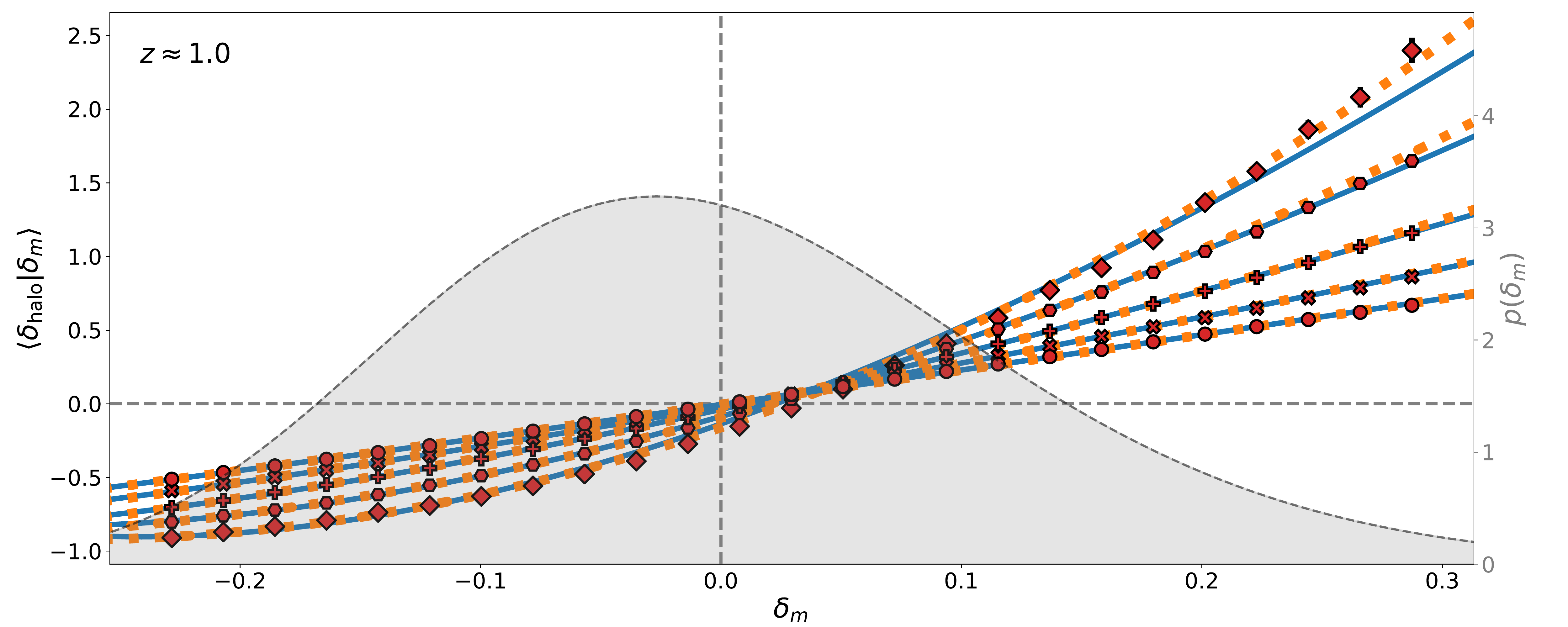}
  \caption{Conditional expectation value $\langle \delta_{\mathrm{halo}}|\delta_m\rangle$ in cylindrical aperture of $R=20$ Mpc$/h$ and $L=150$ Mpc$/h$ at different redshifts and for different halo masses. Mass bins are $M_{200\mathrm{b}}/(10^{13}M_\odot/h) \approx 0.5$ (squares), $\approx 1.0$ (circles), $\approx 2.0$ (crosses), $\approx 4.0$ (pluses), $\approx 8.0$ (hexagons) and $\approx 16.0$ (diamonds). The $z\approx 1.0$ shell of the T17 sims does not resolve the lowest mass bin. The solid blue and dashed orange lines are best-fitting models from 2nd order Lagrangian and Eulerian bias expansions respectively. Errorbars of the symbols are for an all-sky shell and are estimated from a jackknife procedure. The grey shaded area displays the PDF $p(\delta_m)$ (\cf right y-axis) and our binning of $\langle \delta_{\mathrm{halo}}|\delta_m\rangle$ cuts away $2\%$ of the probability from the tails of that PDF.}
  \label{fi:scatter_plots}
\end{figure*}

\begin{table*}
\begin{tabular}{|l|c|c|c|c|c|c|c}
$M_{200\mathrm{b}}/(10^{13}M_\odot/h) \in$ & $b_1^L$ & $b_2^L$ & $\chi_L^2$ & $b_1^E$ & $b_2^E$ & $\chi_E^2$ & $\chi_L^2-\chi_E^2$ \\
\hline
optimally: & & & & & & & \\
$\mathrm{Var}(\chi^2) = 6.78^2$ (statistical) & & & & & & & \\
$\ \ \ \ \ \ \ \ \ \ \ \ \ \ + 2.49^2$ (cov. noise) & & & & & & & \\
$\Rightarrow \chi^2 \sim 23 \pm 7.22$ & & & & & & & \\
\hline
$z\approx 0.5:$ & & & & & & & \\
$[ 0.45 , 0.55 ]$ & 0.39 $\pm$ 0.01 & -0.72 $\pm$ 0.06 & 21.14 & 1.41 $\pm$ 0.01 & -0.62 $\pm$ 0.06 & 20.17 & 0.97 \\
$[ 0.9 , 1.1 ]$ & 0.64 $\pm$ 0.01 & -0.59 $\pm$ 0.08 & 20.28 & 1.66 $\pm$ 0.01 & -0.42 $\pm$ 0.08 & 18.5 & 1.78 \\
$[ 1.8 , 2.2 ]$ & 0.96 $\pm$ 0.02 & -0.43 $\pm$ 0.13 & 14.77 & 1.96 $\pm$ 0.01 & -0.19 $\pm$ 0.14 & 14.13 & 0.64 \\
$[ 3.6 , 4.4 ]$ & 1.42 $\pm$ 0.02 & 0.35 $\pm$ 0.17 & 21.94 & 2.39 $\pm$ 0.02 & 0.73 $\pm$ 0.18 & 23.95 & -1.99 \\
$[ 7.2 , 8.8 ]$ & 2.11 $\pm$ 0.04 & 2.6 $\pm$ 0.28 & 31.19 & 2.99 $\pm$ 0.03 & 3.36 $\pm$ 0.31 & 34.6 & -3.4 \\
$[ 14.4 , 17.6 ]$ & 3.07 $\pm$ 0.06 & 6.72 $\pm$ 0.45 & 26.35 & 3.79 $\pm$ 0.05 & 8.32 $\pm$ 0.51 & 24.78 & 1.58 \\
\hline
$z\approx 0.75:$ & & & & & & & \\
$[ 0.5 , 0.55 ]$ & 0.68 $\pm$ 0.01 & -0.55 $\pm$ 0.12 & 16.53 & 1.7 $\pm$ 0.01 & -0.39 $\pm$ 0.13 & 15.78 & 0.75 \\
$[ 0.9 , 1.1 ]$ & 0.97 $\pm$ 0.01 & -0.26 $\pm$ 0.09 & 22.72 & 1.97 $\pm$ 0.01 & -0.02 $\pm$ 0.09 & 22.05 & 0.67 \\
$[ 1.8 , 2.2 ]$ & 1.41 $\pm$ 0.01 & 0.75 $\pm$ 0.14 & 28.79 & 2.37 $\pm$ 0.01 & 1.11 $\pm$ 0.15 & 37.68 & -8.88 \\
$[ 3.6 , 4.4 ]$ & 2.03 $\pm$ 0.02 & 2.2 $\pm$ 0.19 & 21.84 & 2.95 $\pm$ 0.02 & 2.87 $\pm$ 0.2 & 27.49 & -5.64 \\
$[ 7.2 , 8.8 ]$ & 2.81 $\pm$ 0.04 & 5.16 $\pm$ 0.31 & 21.12 & 3.61 $\pm$ 0.03 & 6.31 $\pm$ 0.33 & 24.2 & -3.07 \\
$[ 14.4 , 17.6 ]$ & 4.08 $\pm$ 0.08 & 11.35 $\pm$ 0.58 & 39.16 & 4.7 $\pm$ 0.06 & 14.25 $\pm$ 0.67 & 22.36 & 16.8 \\
\hline
$z\approx 1.0:$ & & & & & & & \\
$[ 0.9 , 1.1 ]$ & 1.32 $\pm$ 0.01 & 0.29 $\pm$ 0.09 & 35.52 & 2.3 $\pm$ 0.01 & 0.62 $\pm$ 0.1 & 43.46 & -7.93 \\
$[ 1.8 , 2.2 ]$ & 1.86 $\pm$ 0.01 & 1.74 $\pm$ 0.14 & 25.88 & 2.8 $\pm$ 0.01 & 2.28 $\pm$ 0.14 & 41.71 & -15.81 \\
$[ 3.6 , 4.4 ]$ & 2.64 $\pm$ 0.02 & 4.23 $\pm$ 0.26 & 22.58 & 3.52 $\pm$ 0.02 & 5.25 $\pm$ 0.28 & 30.36 & -7.78 \\
$[ 7.2 , 8.8 ]$ & 3.75 $\pm$ 0.05 & 10.19 $\pm$ 0.43 & 29.31 & 4.5 $\pm$ 0.04 & 12.31 $\pm$ 0.48 & 19.76 & 9.54 \\
$[ 14.4 , 17.6 ]$ & 4.96 $\pm$ 0.09 & 16.46 $\pm$ 0.76 & 70.05 & 5.6 $\pm$ 0.07 & 20.6 $\pm$ 0.89 & 45.73 & 24.32 \\
\hline
fitting $\langle \delta_{g}|\delta_m\rangle$ at $z\approx 0.75$ & & & & & & & \\
(with the synthetic galaxies described & & & & & & & \\
in \secref{HOD} and used for \figrefnospace{L_vs_E_Zhai}) & & & & & & & \\
 & 1.77 $\pm$ 0.008 & 2.11 $\pm$ 0.068 & 34.62 & 2.69 $\pm$ 0.007 & 2.69 $\pm$ 0.073 & 59.41 & -24.79 \\
\hline
\end{tabular}
    \caption{Best-fitting parameters and $\chi^2$ values obtained from the fits shown in figures \ref{fi:L_vs_E_Zhai} and \ref{fi:scatter_plots}.}
\label{tab:best_fit_values}
\end{table*}

In \figref{scatter_plots} we show measurements of the conditional expectation value $\langle \delta_{\mathrm{halo}}|\delta_m\rangle$ in three different shells of the T17 simulations (with $z = 0.476,\ 0.751,\ 0.990$) and when averaging halo and matter densities in cylindrical apertures of radius $R=20$ Mpc$/h$ and length $L=150$ Mpc$/h$. The different symbols in the figure represent measurements for different bins of halo mass. We choose bins of $\pm 10\%$ around the central masses $M_{200\mathrm{b}}/(10^{13}M_\odot/h) = 0.5$ (squares), $= 1.0$ (circles), $= 2.0$ (crosses), $= 4.0$ (pluses), $= 8.0$ (hexagons) and $= 16.0$ (diamonds). The $z = 0.99$ shell of T17 does not resolve the lowest of these mass bins and the $z = 0.751$ shell only resolves halos down to exactly $M_{200\mathrm{b}}/(10^{13}M_\odot/h) = 0.5$, i.e. for that shell only the upper half of that bin enters our measurement. For each of the mass bins we measure $\langle \delta_{\mathrm{halo}}|\delta_m\rangle$ in 25 equidistant bins of $\delta_m$ and the lowest and upper most bound of these bins were chosen such as to cut away exactly $2\%$ of the probability from each tail of the underlying matter density PDF $p(\delta_m)$. We estimate the errorbars of each measurement from a jackknife approach \citep{Norberg2009, Friedrich2016}, splitting the all-sky maps of T17 into $196$ sub-patches. The solid blue and dashed orange lines in the figure are best-fitting models from 2nd order Lagrangian and Eulerian bias expansions respectively (\cf \secrefnospace{theory}).

We summarize the best-fitting values of our bias parameters as well as the $\chi^2$ values between best-fitting models and measurements of $\langle \delta_{\mathrm{halo}}|\delta_m\rangle$ in \tabrefnospace{best_fit_values}. Taking into account that the noise in our covariance matrices adds a relative variance of about $\sqrt{2/(196 - 25 - 2)}$ to our best-fitting $\chi^2$ \citep[see \egnospace][we add this noise in quadrature to the expected statistical scatter of $\chi^2$]{Taylor2013}, most of the fits in \figref{scatter_plots} agree with the measurements within either $1\sigma$ or $2\sigma$. For the Lagrangian parametrization, only two of the overall $17$ fits lie outside of $2\sigma$. On average one would expect one such outlier. However, at least one of these outliers is at a very high $\chi^2$ ($6.5 \sigma$) and both of them are at the highest mass of their respective redshifts. Hence, there seems to be a systematic shortcoming of the Lagrangian model for very high halo masses. 

The Eulerian parametrization performs somewhat better in these two instances (though it is still $> 3\sigma$ off for the highest mass bin in the highest redshift shell). But in total, 4 of the Eulerian fits lie outside of $2\sigma$. When ignoring the highest mass bins in the $z=0.75$ and $z=1.0$ shells, the Lagrangian model performs either similarly well or significantly better than the Eulerian one. This is the reason why the Lagrangian model was a significantly better fit to $\langle \delta_{g}|\delta_m\rangle$ for our synthetic galaxy sample discussed in \secref{T17_and_Zhai} (\cf \figref{L_vs_E_Zhai}). The best-fitting parameters and $\chi^2$ of that comparison are also displayed in \tabrefnospace{best_fit_values}.

\begin{figure*}
  \includegraphics[width=0.85\textwidth]{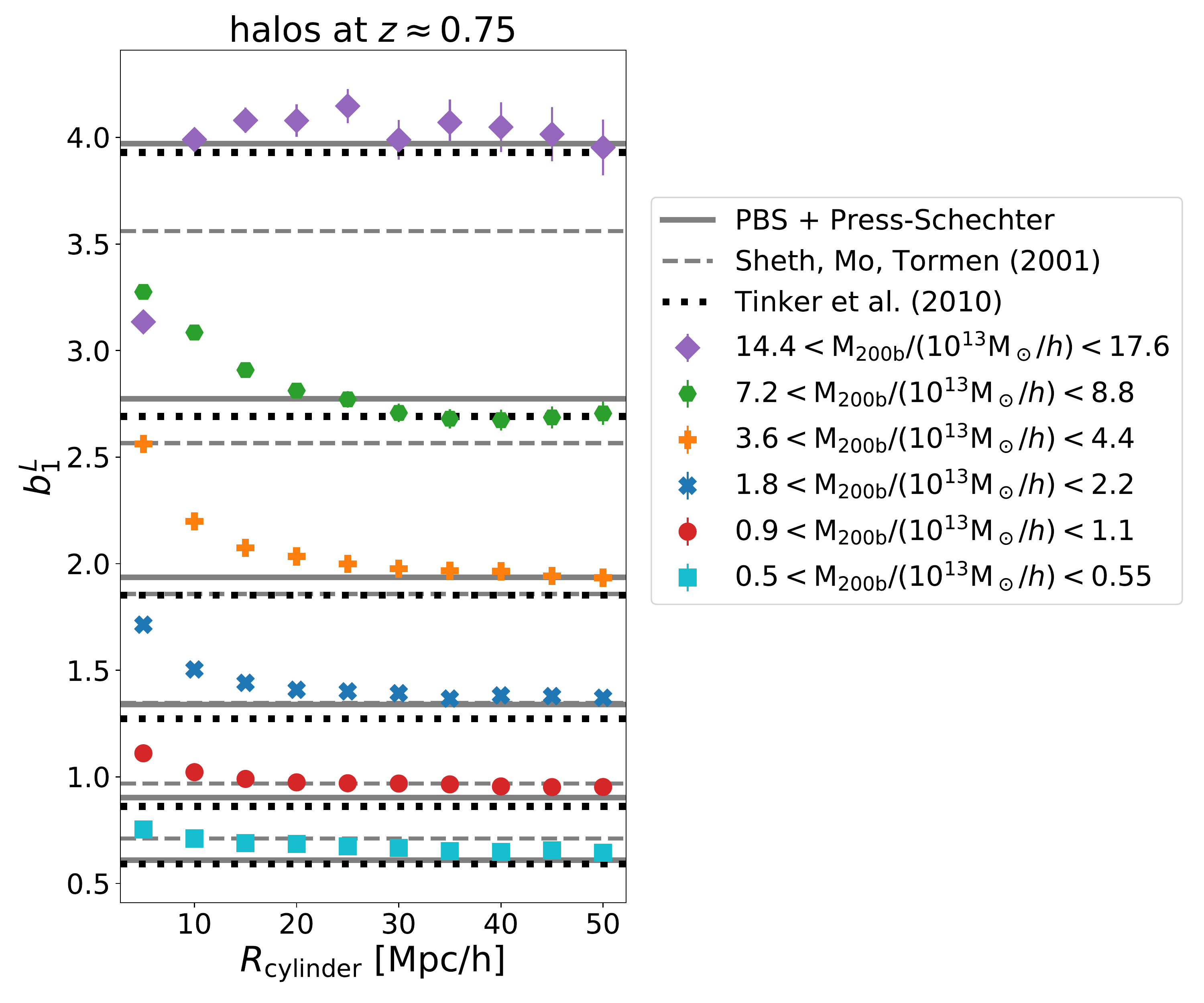}
   \caption{Measurement of linear Lagrangian bias through fits to $\langle \delta_{\mathrm{halo}}|\delta_m\rangle$ in T17 simulated data. Different symbols (and colors) correspond to different halo mass bins and the x-axis represent the radius of the smoothing aperture used to measure $\langle \delta_{\mathrm{halo}}|\delta_m\rangle$. Different horizontal lines correspond to different predictions of $b_1^L(M_{\mathrm{halo}})$ (see main text for details). We chose to leave those lines uncolored for aesthetic reasons.}
  \label{fi:b1L_vs_M_and_R}
\end{figure*}

Let us now investigate whether the best-fitting parameters of our bias models conform to basic theoretical expectations. In \figref{b1L_vs_M_and_R} we show the values obtained for the linear Lagrangian bias $b_1^L$ at $z=0.75$ as a function of the radius of our cylindrical aperture and for all of our different mass bins. The different symbols in the figure show measurements of $b_1^L$ obtained from fitting our Lagrangian model for $\langle \delta_{\mathrm{halo}}|\delta_m\rangle$ to T17 data. Above radii of $R_{\mathrm{cylinder}}\approx 20-30$ Mpc$/h$ there is only a mild scale dependence of these best-fitting values. Different horizontal lines in the figure display different theoretical predictions for the large-scale limit of $b_1^L$. Dashed lines show predictions obtained from the peak-background split (PBS) approach together with Press-Schechter halo mass function \citep{PS1974}, solid lines show predictions based on \citet[][\ie including their \emph{moving barrier} correction]{Sheth2001} and dotted lines show predictions from the fitting formula of \citet{Tinker2010}. All three sets of theoretical predictions match the large scale limit of the bias values we fit with our Lagrangian parametrization to within $10\%$ accuracy. Surprisingly, the Press-Schechter predictions seem to match our measurements of $b_1^L$ best (but with the Tinker et al.\ predictions performing very similarly). At very high masses ($\sim 8\cdot 10^{13} M_\odot/h$ and $\sim 16\cdot 10^{13} M_\odot/h$) we find that the predictions of \citet{Sheth2001} are significantly lower than the other two models the biases measured from $\langle \delta_\mathrm{halo}|\delta_m\rangle$ (and the other sets of predictions).

\begin{figure}
  \includegraphics[width=0.49\textwidth]{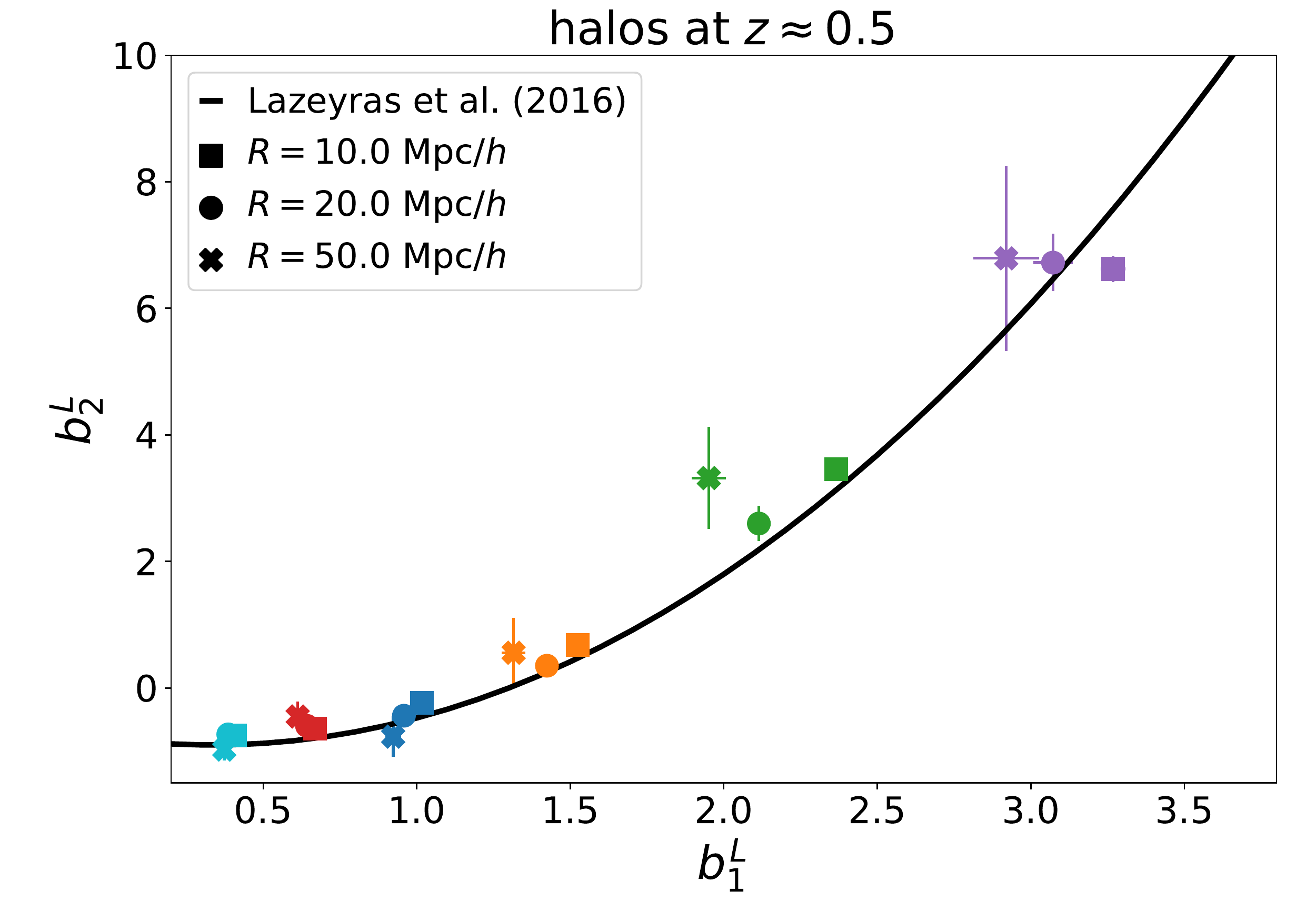}
  
  \includegraphics[width=0.49\textwidth]{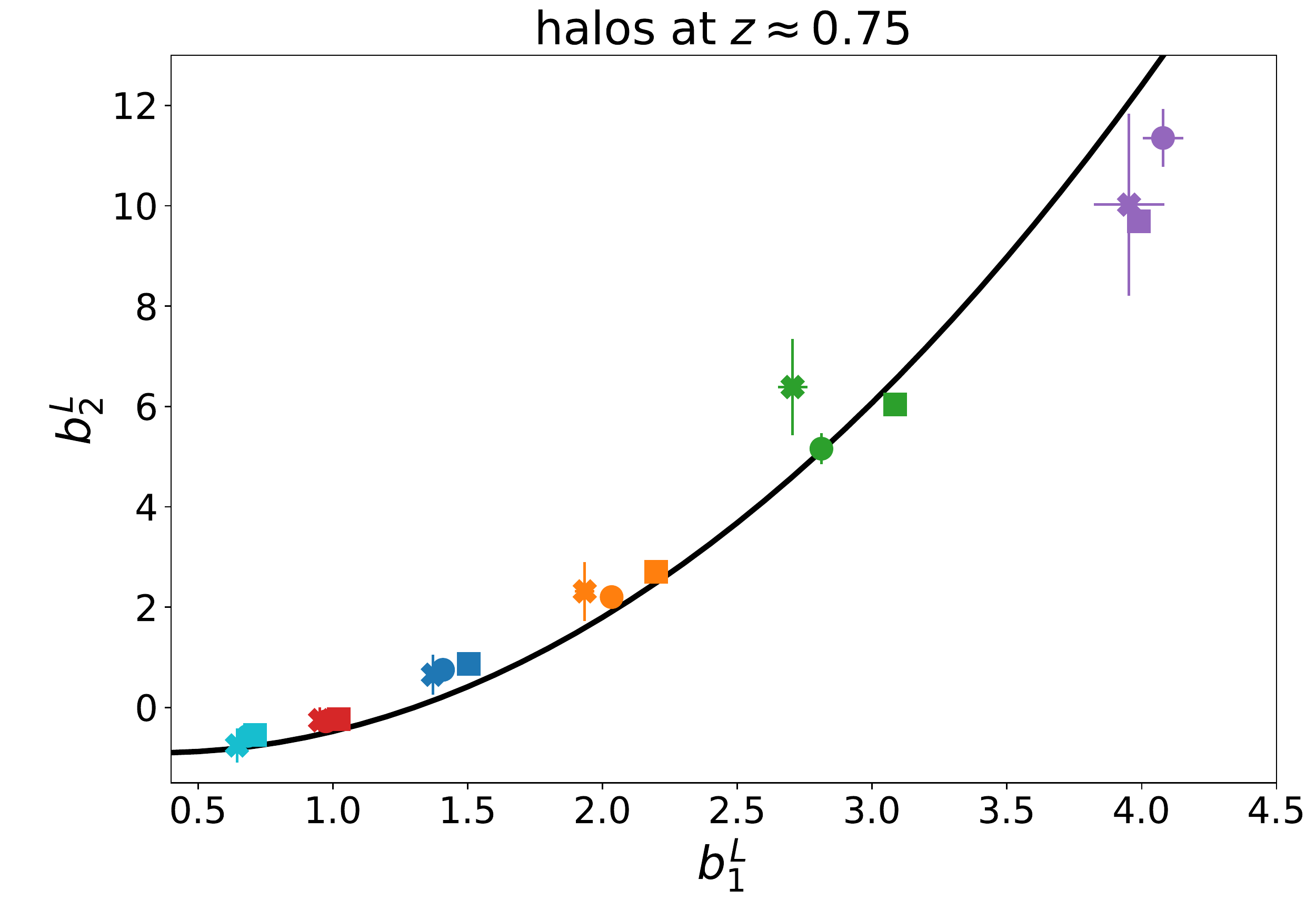}
  
  \includegraphics[width=0.49\textwidth]{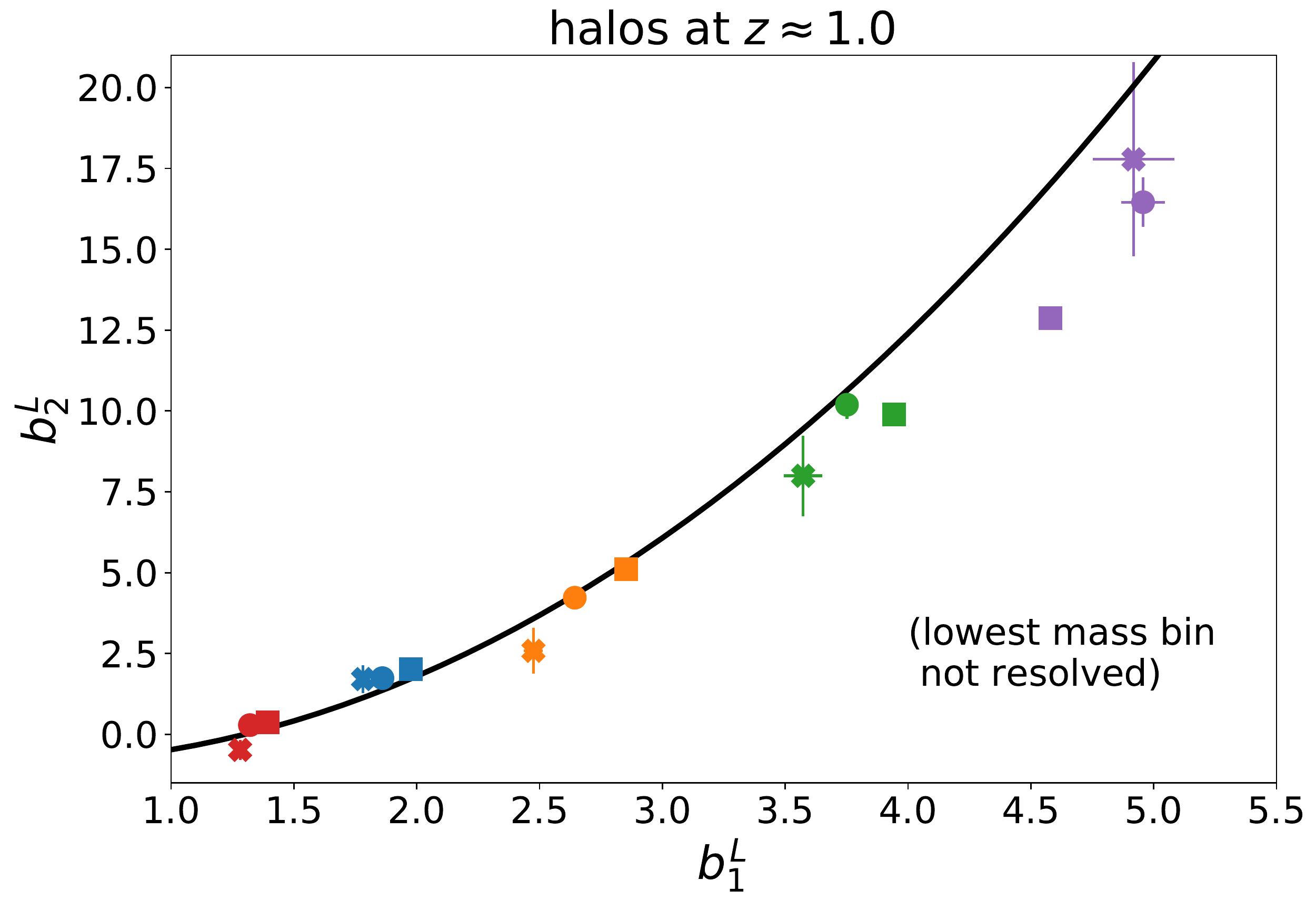}
   \caption{Displaying our measurements of $b_2^L$ from $\langle \delta_{\mathrm{halo}}|\delta_m\rangle$ as a function of the corresponding measurements of $b_1^L$ for different halo mass bins, different radii of our smoothing aperture and in different redshift shells of the T17 sims. The color coding of the mass bins is identical to that of \figref{b1L_vs_M_and_R} (higher bias values correspond to higher masses). The solid lines represent an empirical relation between linear and quadratic bias found by \citet{Lazeyras2016} using a response approach in separate universe simulations (see main text for details).}
  \label{fi:b1L_vs_b2L}
\end{figure}

\citet{Lazeyras2016} have found a tight relationship between linear and quadratic Lagrangian bias, as measured from the response of halo density to changes in the overall matter density in a set of separate universe simulations. We expect our finding to closely match their results, because the expectation values $\langle \delta_\mathrm{halo}|\delta_m\rangle$ resemble exactly that kind of response approach, with each of our apertures representing a (miniature) separate universe. In \figref{b1L_vs_b2L} we show our measurements of $b_1^L$ and $b_2^L$ in the three different redshift shells of the T17 data and for different mass bins. The color coding of the mass bins is identical to that of \figref{b1L_vs_M_and_R} (higher bias values correspond to higher masses) and the different symbols represent fits to $\langle \delta_\mathrm{halo}|\delta_m\rangle$ for different radii of our cylindrical aperture ($R = 10, 20, 50$ Mpc$/h$). The solid line in the figure displays the empirical relation found by \citet{Lazeyras2016}. Despite directly measuring the Lagrangian bias parameters, they present their fit in terms of transformed, Eulerian biases. For reference, we translate that fit to Lagrangian space, which yields
\begin{equation}
\label{eq:Lazeyras_fit}
    b_2^L \approx -0.794 - 0.642\ b_1^L + 0.953\ (b_1^L)^2 + 0.008\ (b_1^L)^2\ .
\end{equation}
This relation indeed closely describes our measurements of $b_1^L$ and $b_2^L$. This is encouraging and confirms that the bias parameters one would measure from our Lagrangian formalism in a PDF-type analysis indeed correspond to the bias parameters that have been investigated in other contexts. This is particularly important when considering combined analysis of the joint PDF $p(\delta_{\mathrm{tracer}},\delta_m)$ and other summary statistics of the cosmic density field. But the agreement seen in \figref{b1L_vs_b2L} does unfortunately not mean that one can hope to eliminate one free parameter from our bias model. The tracers of the cosmic density field available in real analyses are galaxies, and in order to make use of the relation observed in \figref{b1L_vs_b2L} for such analyses one would have to model the Halo-Occupation-Distribution (HOD) of these galaxies, which in itself would introduce a plethora of free parameters \citep[see \egnospace][]{Dvornik2018}. Hence, the strategy we aim for in future data analyses is to fit effective bias parameters for the tracer samples at hand, as we have \eg done in \secrefnospace{T17_and_Zhai}. This is also why we do not further pursue accurate modelling of $b_1^L$ as a function of halo mass.

\begin{figure}
  \includegraphics[width=0.49\textwidth]{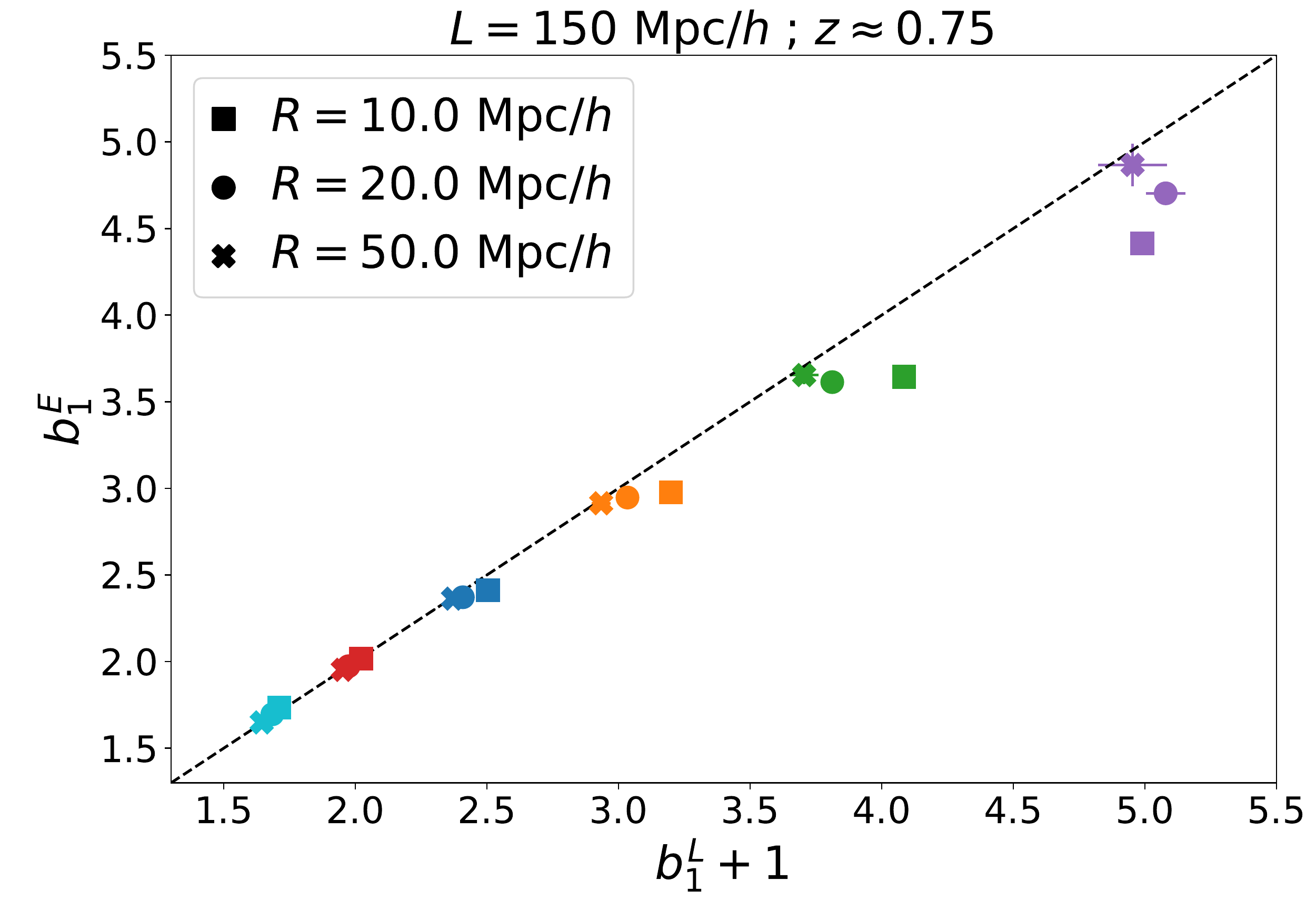}
  
  \includegraphics[width=0.49\textwidth]{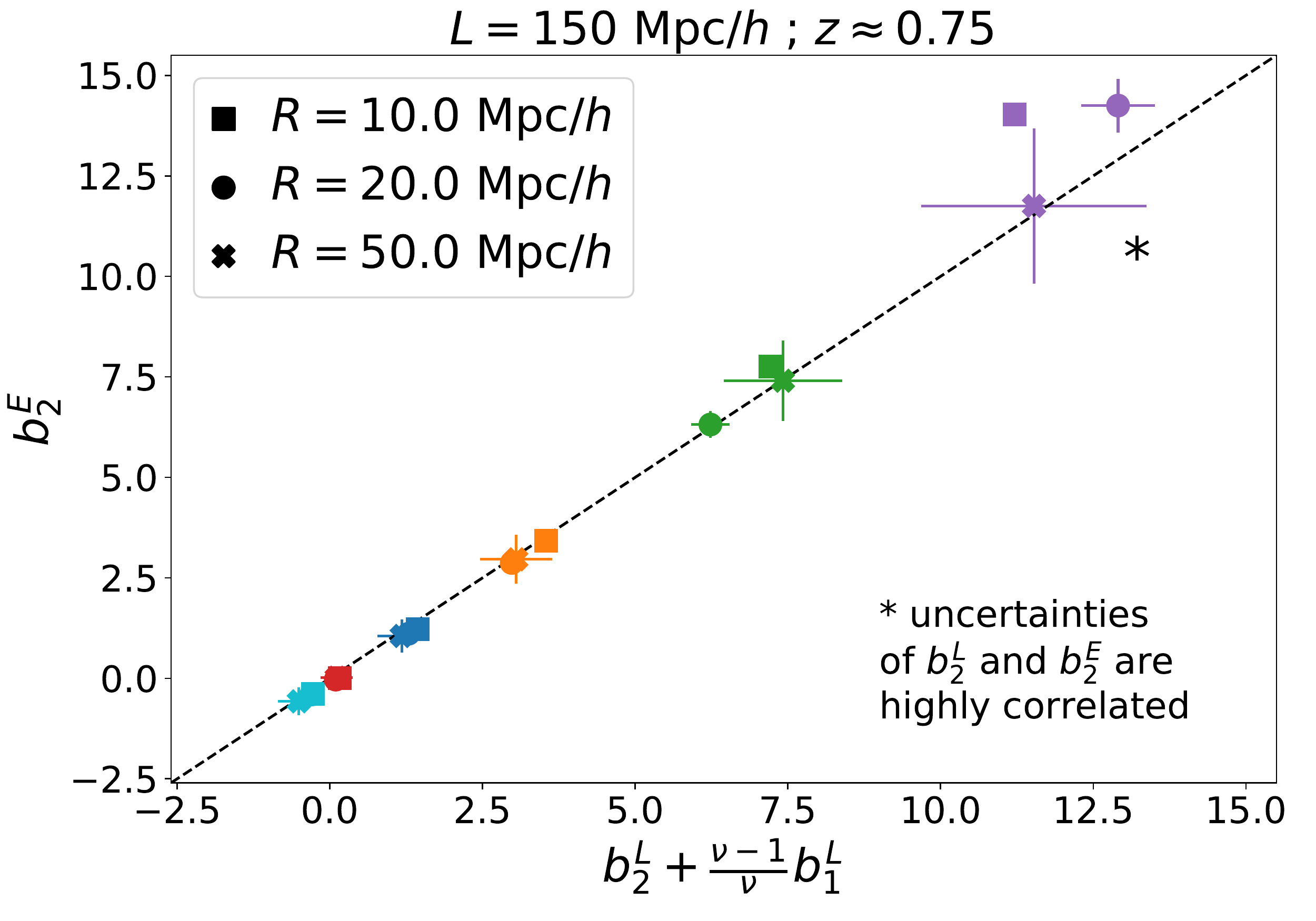}
   \caption{Testing whether standard relations between Lagrangian and Eulerian bias coefficients hold for our measurements of these coefficients from $\langle \delta_{\mathrm{halo}}|\delta_m\rangle$ in the large scale limit (see main text for details). Different colors again represent different halo mass bins and the color coding is the same as that in \figref{b1L_vs_M_and_R}. In the lower panel, the agreement between Eulerian and Lagrangian parameters looks suspiciously good, given the statistical uncertainties of our fits. This is caused by the fact that the measurement uncertainties for both sets of parameters are highly correlated.}
  \label{fi:bL_vs_bE}
\end{figure}

In a next step, we want to check for consistency between the bias parameters measured from our Lagrangian and Eulerian models for $\langle \delta_\mathrm{halo}|\delta_m\rangle$. In the large-scale limit $b_1^L, b_2^L$ and $b_1^E, b_2^E$ should be related by
\begin{align}
\label{eq:bE_as_function_of_bL}
    b_1^E \approx&\ 1+b_1^L \\
    b_2^E \approx&\ \frac{\nu-1}{\nu} b_1^L + b_2^L\ ,
\end{align}
where typically one assumes $\nu = 21/13$ \citep{Wagner2015, Lazeyras2016}. The derivation of that value uses a spherical collapse approximation which may not be appropriate for our cylindrical apertures. Substituting spherical with cylindrical collapse one arrives at $\nu = 7/5$ \citep{Uhlemann2018c}. In our situation we find both values for $\nu$ to give very similar values of $b_2^E$ (as calculated from $b_1^L$ and $b_2^L$) and for cylinders of finite length, the truth is anyway expected to lie between both choices \citep[see again][]{Uhlemann2018c}. So in the following we will stick with the spherical value such that $(\nu - 1)/\nu = 8/21$. In \figref{bL_vs_bE} we plot our measurements of $b_1^E$ as a function of $1+b_1^L$ and our measurements of $b_2^E$ as function of $(\frac{8}{21} b_1^L + b_2^L)$. Different colors again represent different halo mass bins and different symbols represent different aperture radii. One can see that for our largest aperture ($50$ Mpc/$h$) the agreement with the relations \ref{eq:bE_as_function_of_bL} is indeed excellent (note that the measurement uncertainties of $b_2^E$ and $b_2^L$ are highly correlated, which is the reason why the measurements in the bottom panel are suspiciously spot on). This demonstrates that the machinery we have developed in \secref{theory} indeed represents a sensible Lagrangian bias model for PDF statistics.

Finally, we want to compare our linear bias values obtained from $\langle \delta_\mathrm{halo}|\delta_m\rangle$ to the halo biases that would be inferred from measurements of large-scale 2-point statistics \citep[\cfnospace][who have performed an analogous study for spherical apertures in simulation snapshots]{Manera2011}. Since we are using the T17 data in radial shells, we will consider the angular power spectra of the matter density and halo density fields in these shells projected onto the sky. Let $C_\ell^{mm}$ be the auto power spectrum of the matter density field in a particular shell, and let $C_\ell^{hm}$ be the cross power spectrum of matter and halo density. Following a similar procedure to that of \citet{Lazeyras2016} we assert that those are related by
\begin{equation}
\label{eq:Lazeyras_2pt_fit}
    C_\ell^{hm} \approx (b_1^{2\mathrm{pt}} + b_{\mathrm{NL}}^{2\mathrm{pt}}\ \ell^2)\ C_\ell^{mm}\ .
\end{equation}
Here $b_1^{2\mathrm{pt}}$ is the linear Eulerian halo bias (in the 2-point function context) and the term proportional to $b_{\mathrm{NL}}^{2\mathrm{pt}}\ \ell^2$ aims to capture corrections from non-linear (\resp scale dependent) bias. We fit the above relation to measurements of $C_\ell^{hm}$ and $C_\ell^{mm}$ in T17 data. This has the advantage that we do not need to employ any analytic modelling of the involved power spectra. Also, it removes the dependence of the fit on shot-noise. We follow \citet{Lazeyras2016} in restricting the fit to co-moving wave numbers below $k=0.06\ h/$Mpc. Coincidentally, this roughly corresponds to real space scales of $\pi/k \gtrsim 50$ Mpc$/h$, \ie to about the largest aperture radius in which we have measured $\langle \delta_\mathrm{halo}|\delta_m\rangle$. If $w$ is the average co-moving distance of a shell, then the co-moving wave number $k$ probed by an angular mode $\ell$ is approximately $\ell/w$. Hence, we restrict ourselves to modes $\ell \leq 0.06 \cdot w\ h/$Mpc.

To estimate the statistical uncertainties of this fit, let us assume that bias is perfectly linear, and that both the matter density and halo density field are Gaussian random fields. These assumptions are likely sufficient for our two-point analysis, since the power spectrum covariance at small scales (where the assumptions may break down) will be dominated by shot-noise \citep[\cfnospace][]{Friedrich2020b}. This will especially be the case for the narrow bins in halo mass that we consider here.

\begin{figure}
  \includegraphics[width=0.49\textwidth]{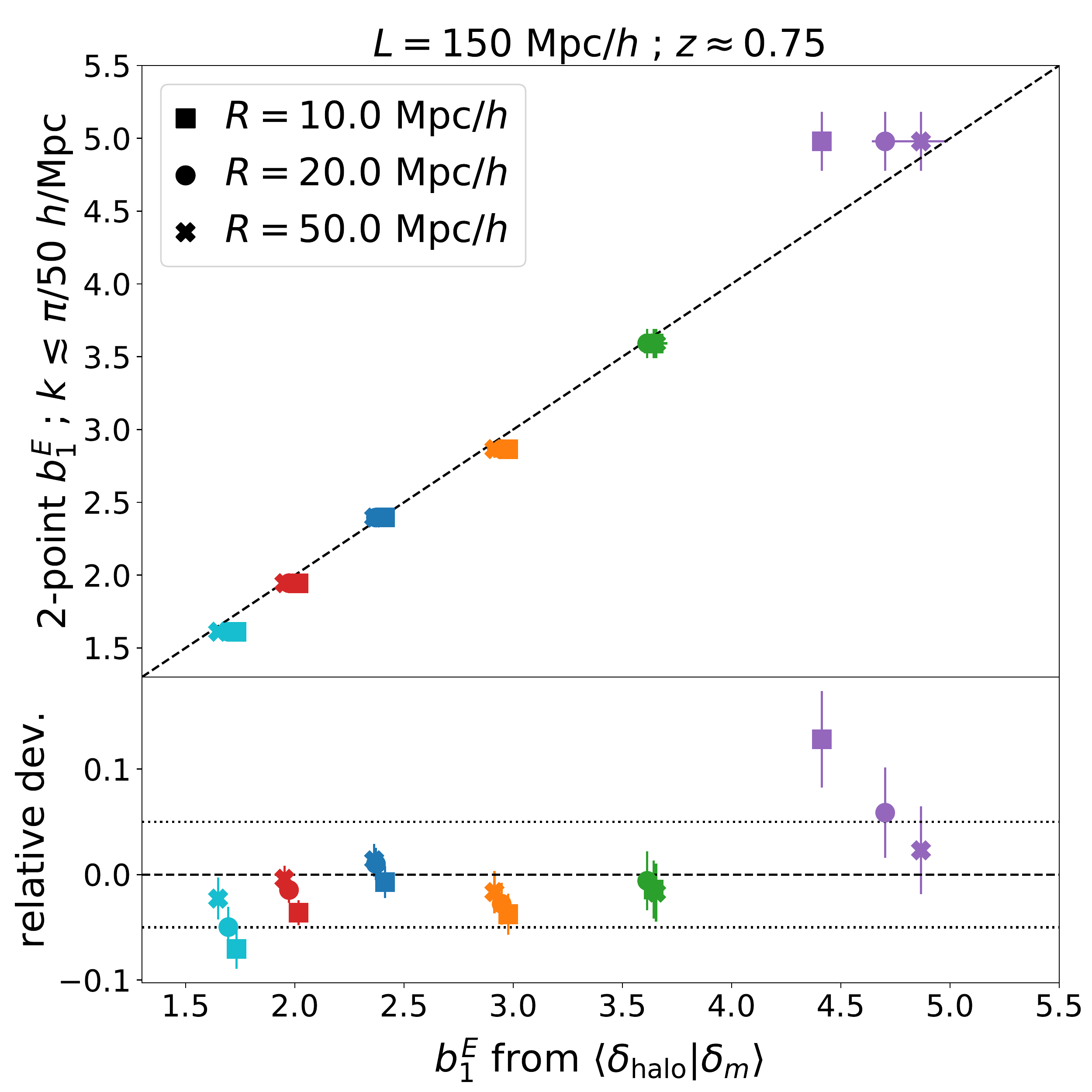}
   \caption{Testing whether the Eulerian linear bias $b_1^E$ measured from $\langle \delta_{\mathrm{halo}}|\delta_m\rangle$ agrees with the bias measured from comparing the auto power spectrum of matter density fluctuations to the cross power spectrum of matter and halo density fluctuations (see main text for details, the lower panel shows relative deviations between the two sets of measurements). Different colors again represent different halo mass bins (\cf \figref{b1L_vs_M_and_R} for the color coding). To perform the power spectrum fits we only considered scales with $\pi / k \gtrsim 50$ Mpc/$h$.}
  \label{fi:2pt_bias_vs_PDF_bias}
\end{figure}

A measurement of the matter auto power spectrum will be given by
\begin{equation}
    \hat C_\ell^{mm} = \frac{1}{2\ell + 1} \sum_{M=-\ell}^\ell |a_{\ell M}|^2
\end{equation}
where $a_{\ell M}$ are the spherical harmonics coefficients of the matter density field projected onto the sky. Similarly, a measurement of $C_\ell^{hm}$ will be given by
\begin{equation}
    \hat C_\ell^{hm} = \frac{1}{2\ell + 1} \sum_{M=-\ell}^\ell a_{\ell M}^*\ (b_{\ell M} + \epsilon_{\ell M})
\end{equation}
where $\epsilon_{\ell M}$ represents shot-noise and $b_{\ell M}$ are the spherical harmonics coefficients of the (hypothetical) shot-noise free halo density field. We need to know the covariance matrix of
\begin{align}
\hat C_\ell^{hm} - b_1^{2\mathrm{pt}}\hat C_\ell^{mm} \approx \frac{1}{2\ell+1}\sum_{M=-\ell}^\ell a_{\ell M}^*\ \epsilon_{\ell M}\ .
\end{align}
Within our Gaussianity and linearity assumption it is easy to see that this covariance is diagonal and that the variances for each value of $\ell$ are given by
\begin{align}
    \mathrm{Var}\left(\hat C_\ell^{hm} - b_1^{2\mathrm{pt}}\hat C_\ell^{mm}\right) = \frac{C_\ell^{mm}}{(2\ell +1)n_{\mathrm{halo}}} \approx \frac{\hat C_\ell^{mm}}{(2\ell +1)n_{\mathrm{halo}}}\ .
\end{align}
Here $n_{\mathrm{halo}}$ is the number density of halos (projected onto the sky) and we have assumed that the shot-noise is uncorrelated to the underlying matter density field. So the figure of merit that we are optimising in order to fit for the bias parameters in \eqnref{Lazeyras_2pt_fit} is 
\begin{align}
     &\ \chi^2[b_1^{2\mathrm{pt}}, b_{\mathrm{NL}}^{2\mathrm{pt}}]
    \approx \nonumber \\
     &\ \sum_{\ell < \ell_{\max}} \frac{\left(\hat C_\ell^{hm} - (b_1^{2\mathrm{pt}} + b_{\mathrm{NL}}^{2\mathrm{pt}}\ \ell^2)\ \hat C_\ell^{mm}\right)^2}{\hat C_\ell^{mm}}\ (2\ell +1)n_{\mathrm{halo}} \ .
\end{align}
The best-fitting reduced $\chi^2$ values we obtain this way indeed scatter closely around 1. In \figref{2pt_bias_vs_PDF_bias} we display the corresponding best-fitting values of $b_1^{2\mathrm{pt}}$ as a function of $b_1^{E}$ obtained from the conditional expectation value $\langle \delta_\mathrm{halo}|\delta_m\rangle$ in different halo mass bins (the same bins and color coding as before). Different symbols again correspond to different aperture radii. One can indeed see that the two types of bias measurements agree in the large-scale limit. For our largest aperture radius, the relative agreement is better than $3\%$ in all mass bins and within the statistical uncertainties of the 2-point fit. The systematic shift of $b_1^{E}$ when going to smaller radii does not necessarily signify a systematic difference between 2-point function and PDF biases but rather implies a general scale-dependence of bias when moving to smaller scales. Note especially, that we only measured our power spectra on scales of $\pi/k \gtrsim 50$ Mpc$/h$. Allowing the 2-point fit to use even smaller scales leads to a shift in $b_1^{\mathrm{2pt}}$ similar to that observed in $\langle \delta_\mathrm{halo}|\delta_m\rangle$.

\citet{Lazeyras2016} perform a similar test also for the quadratic bias coefficients. This would require us to either model the non-linear part of the halo power spectra or to measure complicated combinations of bispectra in the T17 shell. We do not attempt that because we take our comparison for the linear coefficients in combination with the results obtained for the quadratic coefficients in Figures \ref{fi:b1L_vs_b2L} and \ref{fi:bL_vs_bE} as sufficient indication that our language for quadratic bias in the PDF agrees with the parametrizations that appear in more standard contexts.

\subsection{Shot-noise of halos and galaxies}
\label{sec:shot-noise_details}

\begin{figure}
  \includegraphics[width=0.47\textwidth]{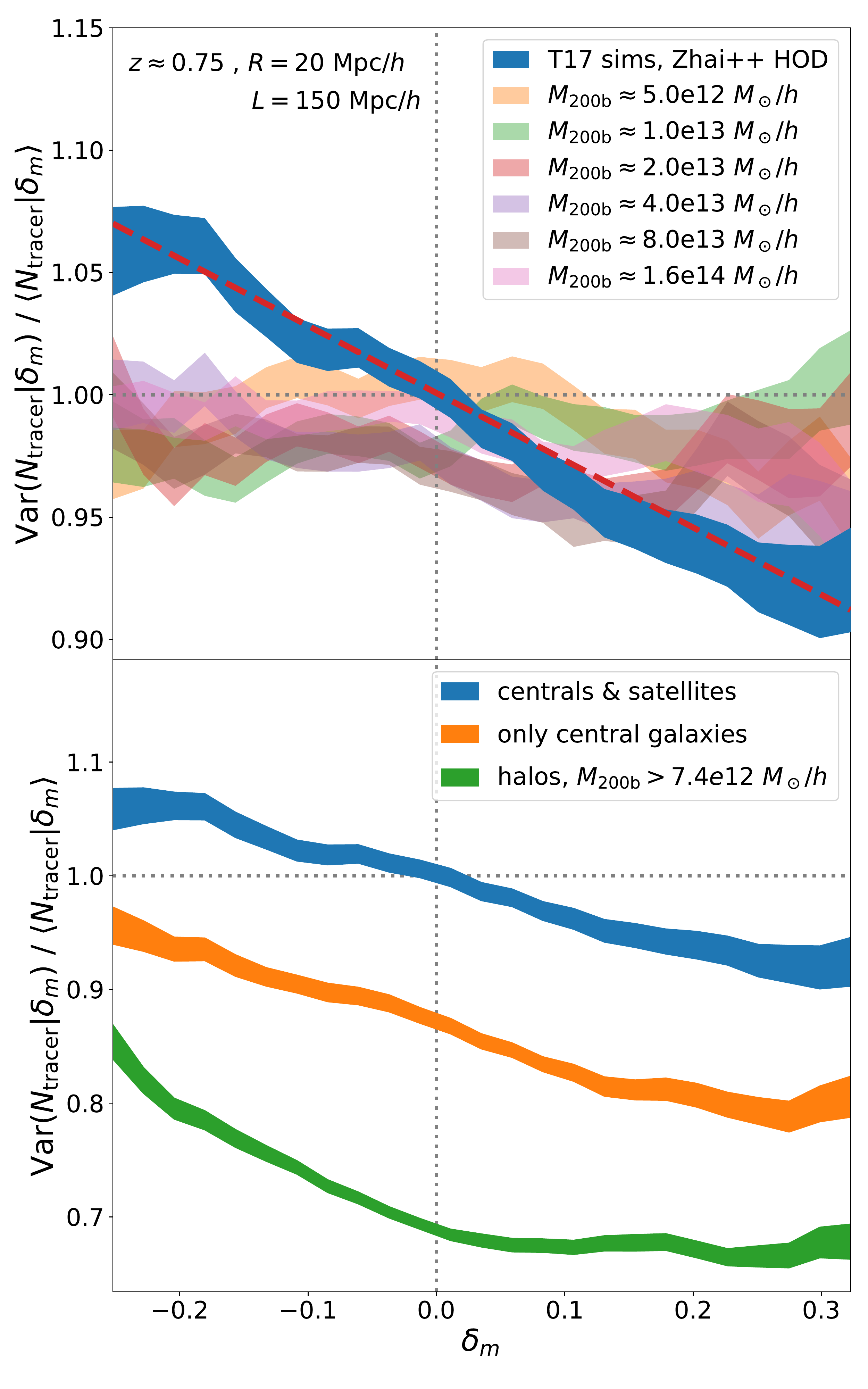}
  \caption{Upper panel: Ratio between the variance and expectation value of tracer counts in cylindrical apertures as a function of the matter density contrast in those apertures. For Poisson shot-noise this ration should be equal to $1$. The red dashed line represents a linear fit to the ratio observed for our synthetic galaxy sample in the T17 shell. Lower panel: same ratio but considering different tracer samples.}
  \label{fi:shot-noise}
\end{figure}

We conclude this section by investigating the shot-noise of our different tracer samples in more detail. The upper panel of \figref{shot-noise} plots the ratio $\mathrm{Var}(N_{\mathrm{tracer}}|\delta_m)/\langle N_{\mathrm{tracer}}|\delta_m \rangle$ measured in the T17 shell with $z\approx 0.75$ and with cylindrical apertures of radius $R=20$ Mpc$/h$. The bins in $\delta_m$ are the same as those we have considered for $\langle \delta_{\mathrm{tracer}}|\delta_m \rangle$ in the previous subsections and the statistical uncertainties have been estimated using the same jackknife procedure as before. The dark blue band in the figure represents the ratio measured for the synthetic galaxy sample described in \secref{HOD} while the semi-transparent bands represent the same halo mass bins as considered previously. For Poissonian shot-noise, the ratios $\mathrm{Var}(N_{\mathrm{tracer}}|\delta_m)/\langle N_{\mathrm{tracer}}|\delta_m \rangle$ should be equal to $1$. For the different halo mass bins it is slightly below that, with the variances of halo counts being on average about $3\%$ below the Poisson value and with a slight increase of this effect towards higher matter densities. For our synthetic galaxies the situation is quite different: they show variances that are up to $8\%$ above the Poissonian value for negative $\delta_m$, which then steeply fall to give sub-Poissonian variances for positive $\delta_m$. Our HOD prescription of \secref{HOD} should in principle return a weighted average over halos of different masses and at a first glance it is surprising that this would give such a qualitatively and quantitatively different behavior of shot-noise compared to the individual mass bins.

\begin{figure}
  \includegraphics[width=0.47\textwidth]{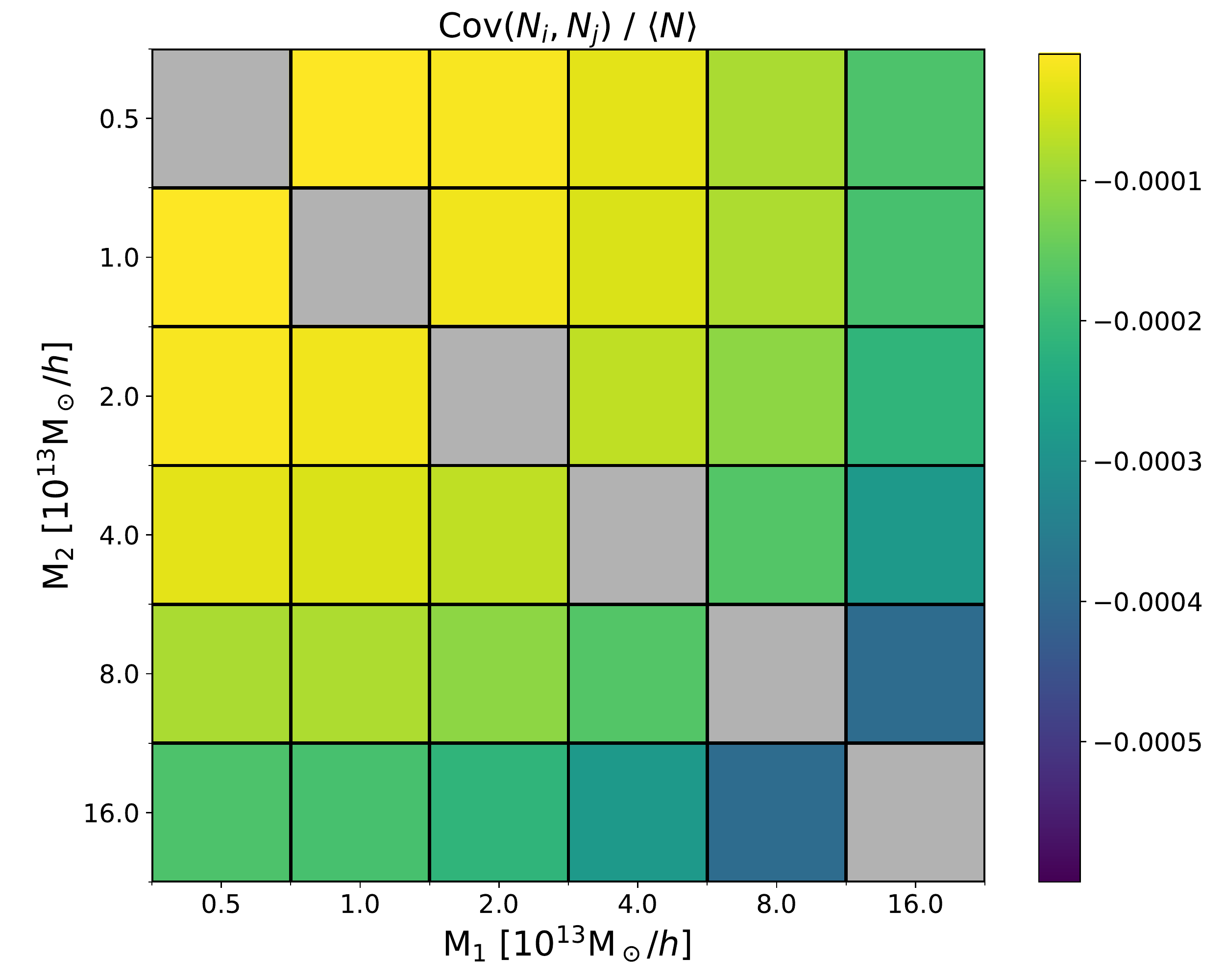}
  
  \includegraphics[width=0.47\textwidth]{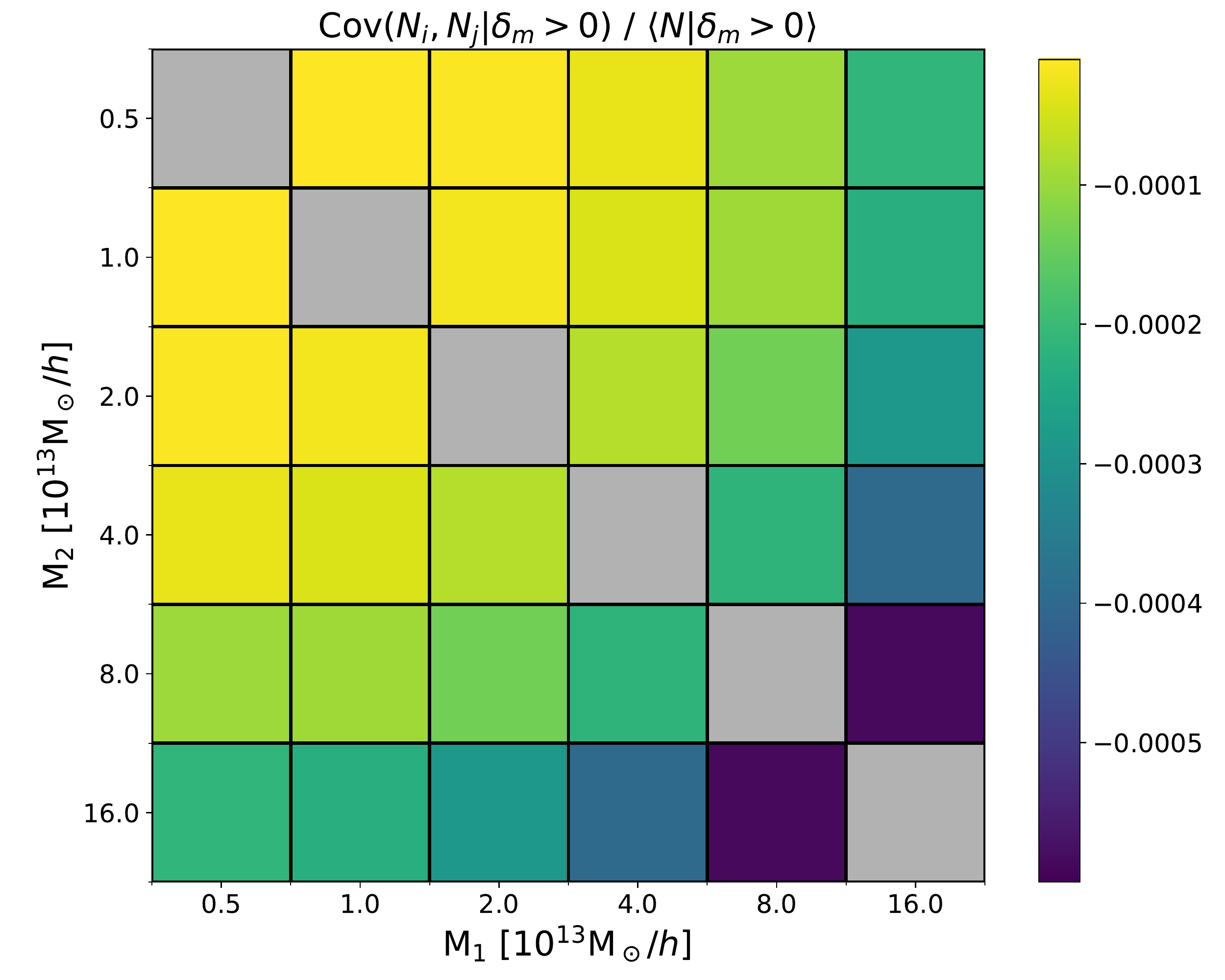}
  
  \includegraphics[width=0.47\textwidth]{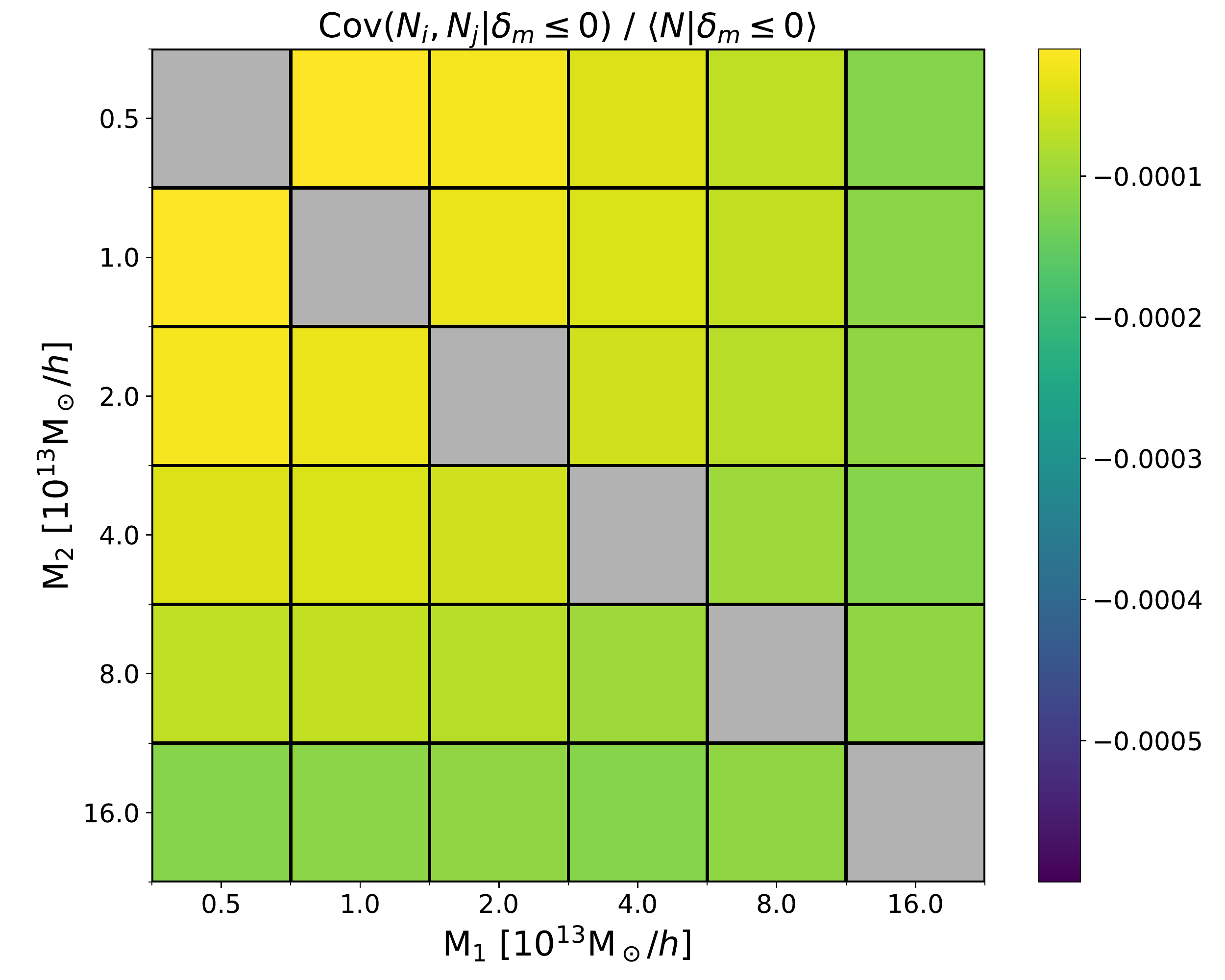}
   \caption{The covariance matrix of shot-noise-only maps for halos in different mass bins, divided by the mean number count of all halos (\ie the matrix $\mathrm{Cov}(N_i, N_j)/\langle N \rangle $ appearing on the right hand-side of \eqnrefnospace{shot-noise_covariance}). The figure uses our fiducial aperture of $R_\mathrm{cyl} = 20$ Mpc$/h$ and the $z\approx 0.75$ shell of our T17 data. The upper panel measures the correlations from the full T17 shell, the middle panel only from parts of the shell where $\delta_m > 0$ and the lower panel from parts where $\delta_m \leq 0$.}
  \label{fi:shot-noise_correlation}
\end{figure}

To understand this in more detail, note that for halo bins with a very narrow mass range the ratio $\mathrm{Var}(N|\delta_m)/\langle N|\delta_m\rangle$ will always tend to $1$. This is because for a small enough mass range, there will always be either 0 or 1 halo in any of our apertures. And in that case, the shot-noise becomes binomial with number of trials $N_\mathrm{trial}= 1$ and probability of failure $1-p \approx 1$, which results in $\mathrm{Var}(N|\delta_m)/\langle N|\delta_m\rangle \approx 1$. Now what happens to the ratio when summing over many of these narrow bins? Let $N_i,\ i=1,\dots,n$ be the counts of $n$ of such narrow halo bins in our aperture and consider their sum
\begin{equation}
    N = \sum_{i=1}^n N_i\ .
\end{equation}
Obviously, the expectation value of $N$ is just the sum of the expectation values of the $N_i$,
\begin{equation}
    \langle N | \delta_m \rangle = \sum_{i=1}^n \langle N_i | \delta_m \rangle\ ,
\end{equation}
where we have inserted a dependence on $\delta_m$ to be closer to our situation of interest. For the variance of $N$ the situation is more complicated since
\begin{align}
\label{eq:shot-noise_covariance}
    \mathrm{Var}(N | \delta_m) =&\ \sum_{i,j} \mathrm{Cov}(N_i, N_j | \delta_m)\nonumber \\
    \approx &\ \sum_{i} \langle N_i| \delta_m \rangle + \sum_{i \neq j} \mathrm{Cov}(N_i, N_j | \delta_m)\nonumber \\
    \Rightarrow \frac{\mathrm{Var}(N | \delta_m)}{\langle N | \delta_m \rangle} \approx&\ 1 + \frac{\sum_{i \neq j} \mathrm{Cov}(N_i, N_j | \delta_m)}{\langle N | \delta_m \rangle}\ .
\end{align}
Now each of the finite mass bins in the upper panel of \figref{shot-noise} can be seen as a sum over many, even narrower mass bins. In order for the shot-noise of the finite mass bins to be sub-Poisson we would hence need the covariance between the narrow bins to be negative. In the following intuitive sense this would indeed be expected: if there are already a lot of halos of one bin in our aperture, one would expect less mass to be left for forming other halos which would cause negative correlations among the shot-noise of the two mass bins. This is also in line with arguments of halo-exclusion \citep[\egnospace][]{Baldauf2013, baldauf21} and with the finding that certain weighting schemes among halo masses can reduce tracer stochasticity \citep[\egnospace][]{Hamaus2010, Jee2012, Uhlemann2018a}. We could in principle estimate the covariance $\mathrm{Cov}(N_i, N_j | \delta_m)$ from our simulated data. Unfortunately, for very narrow mass bins such an estimate will be extremely noisy, because the standard deviation of off-diagonal elements of the estimate will be proportional to diagonal elements of the covariance \citep{Taylor2013} which are significantly higher than the off-diagonal elements in the limit of narrow bins. Nevertheless, to qualitatively test our above considerations, we measure the covariance of the shot-noise of wide mass bins instead. We choose those to be centred around the same masses as our previous bins, but widen the mass ranges to touch each other (but we keep the binning logarithmic). We then re-fit the Eulerian bias model to these new bins and apply the best-fitting parameters to the dark matter density field $\delta_m$ in the T17 data. This way we effectively obtain a shot-noise free estimate of the halo density field which we can then subtract from the actual halo density field to obtain shot-noise-only maps. 

In the upper panel of \figref{shot-noise_correlation} we show the covariance matrix of these shot-noise-only maps, divided by the mean number count of all halos (\ie the matrix $\mathrm{Cov}(N_i, N_j)/\langle N \rangle $ appearing on the right hand-side of \eqnrefnospace{shot-noise_covariance}), using again the T17 shell at $z\approx 0.75$ and filtering with $R_\mathrm{cyl} = 20$ Mpc$/h$. All off-diagonal elements of this matrix are indeed negative! We can furthermore split the T17 shell into regions of positive and negative $\delta_m$. The middle panel of \figref{shot-noise_correlation} shows $\mathrm{Cov}(N_i, N_j)/\langle N \rangle $ obtained only from overdense regions while the lower panel uses only underdense regions. Most of the elements of $\mathrm{Cov}(N_i, N_j)/\langle N \rangle $ are more negative for $\delta_m > 0$ than they are for $\delta_m < 0$. From this behaviour of the shot-noise correlation matrix we can draw the following qualitative conclusions: We expect the shot-noise of halos with a wide mass range to be even more sub-Poissonian than what we observed for our narrow mass bins in the upper panel of \figref{shot-noise}. And we expect the shot-noise of wide halo bins to be more sub-Poissonian in overdense regions than in underdense regions.

\begin{figure}
  \includegraphics[width=0.47\textwidth]{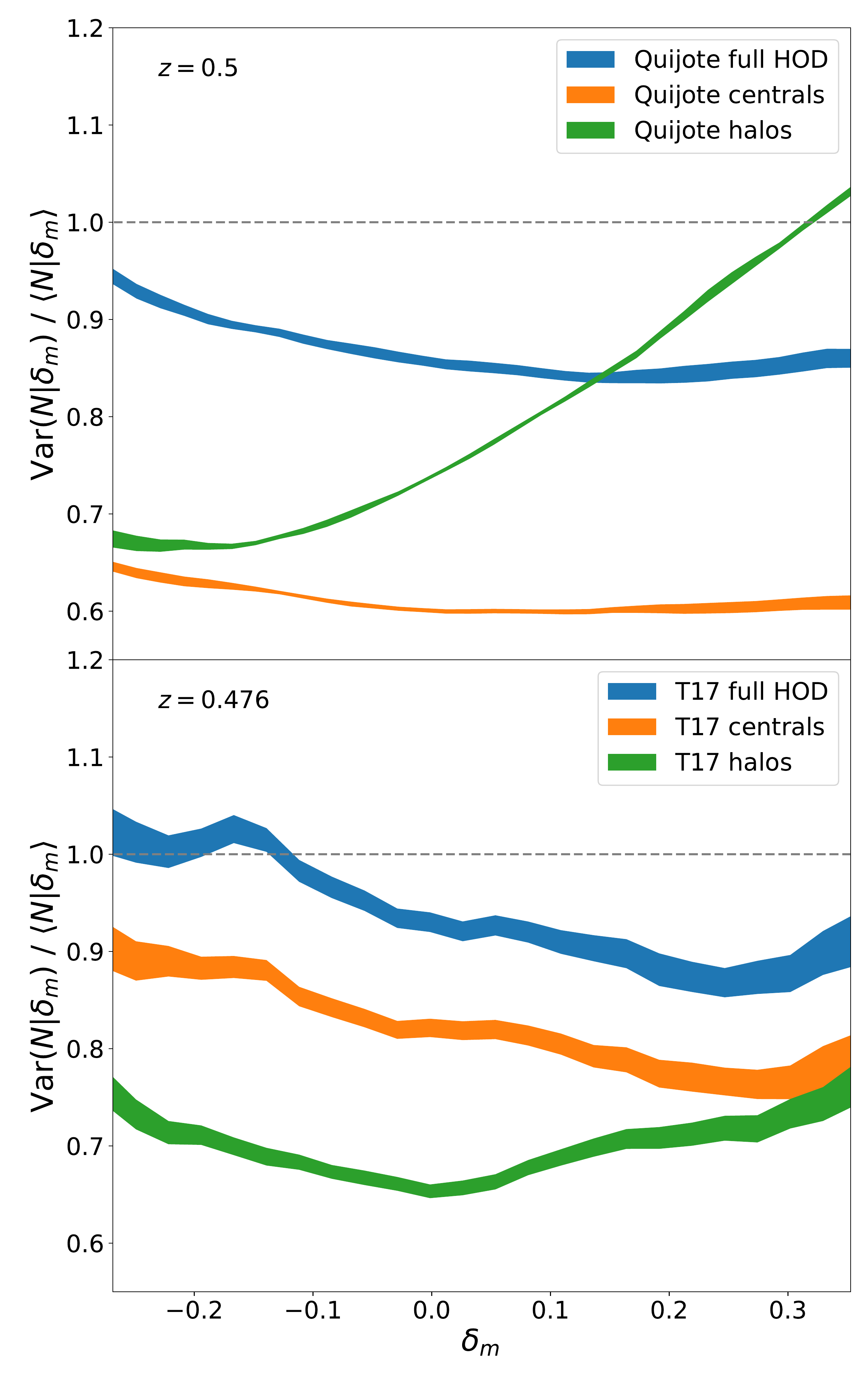}

   \caption{Same as the lower panel of \figref{shot-noise} but at $z = 0.5$ in the Quijote simulations (upper panel) and $z\approx 0.476$ in our T17 mock galaxy catalog (lower panel).}
  \label{fi:Quijote_vs_T17}
\end{figure}

These qualitative statements are indeed confirmed by the green band in the lower panel of \figrefnospace{shot-noise}, which shows the ratio $\mathrm{Var}(N_{\mathrm{tracer}})/\langle N_{\mathrm{tracer}} \rangle$ for a halo mass bin that includes all halos that enter our HOD as described in \secrefnospace{HOD} (\ie all halos with masses $M_{200\mathrm{b}}> 7.4\cdot 10^{12} M_\odot/h$). The shot-noise of that pure halo sample is strongly sub-Poissonian and the ratio $\mathrm{Var}(N_{\mathrm{tracer}})/\langle N_{\mathrm{tracer}} \rangle$ is lower in overdense regions than it is in underdense regions. In that panel, we also plot the shot-noise behaviour of the central galaxies within our mock galaxy sample (orange band; \ie those galaxies that are central to their host halo, \cf \secrefnospace{HOD}) as well as the behaviour of the full synthetic galaxy sample (blue band; same as in upper panel). These bands show a subsequent increase in the ratio $\mathrm{Var}(N_{\mathrm{tracer}})/\langle N_{\mathrm{tracer}} \rangle$ with centrals being already less sub-Poisson than the halos and satellites showing almost Poissonian noise again. One could think that this suggest that the randomness in the HOD is increasing the shot-noise \wrt a pure halo sample and hence pushes the noise closer to Poisson again (or even beyond). However, the situation is more complicated as we explain in the following.

The behaviour of shot-noise in our mock LRG sample strongly differs from what has been observed in a different mock sample by \citet{Friedrich2018} or even for real DES galaxies by \citet{Gruen2018}. They have observed super-Poissonian noise for redMaGiC(-like) galaxies \citep{Rozo2016} that increases with increasing matter density. Since this is so different from our findings, we want to cross check the latter \wrt data from a different N-body simulation - the Quijote suite (\cf \secrefnospace{Quijote}). For that suite we only have a snapshot available at $z=0.5$. So in order to compare our Quijote results to the T17 results we also repeat some of our measurements in the T17 shell at $z\approx 0.476$. We populate halos in both of these data sets with the same HOD as before \citep[see also][for the general methodology]{Hahn2021} and we again consider the ratio $\mathrm{Var}(N_{\mathrm{tracer}})/\langle N_{\mathrm{tracer}} \rangle$ for halos (with $M_{200\mathrm{b}}> 7.4 \cdot 10^{12} M_\odot/h$), for central galaxies and for the full mock galaxy samples. This is not entirely realistic, since the HOD description of \citet{Zhai2017} has been specifically fit to LRGs at $z\approx 0.6-0.9$, but it should nevertheless suffice for a qualitative comparison. Note also that we have only had access to $M_{\mathrm{vir}}$ for the Quijote halos, instead of $M_{200\mathrm{b}}$. But we find that a lower mass cut at $M_{\mathrm{vir}} = 6.986 \cdot 10^{12}M_\odot/h$ within the T17 sims gives a similar halo density as the cut in $M_{200\mathrm{b}}$, so we apply this $M_{\mathrm{vir}}$ cut in Quijote.

\begin{figure}
\begin{center}
  \includegraphics[width=0.4\textwidth]{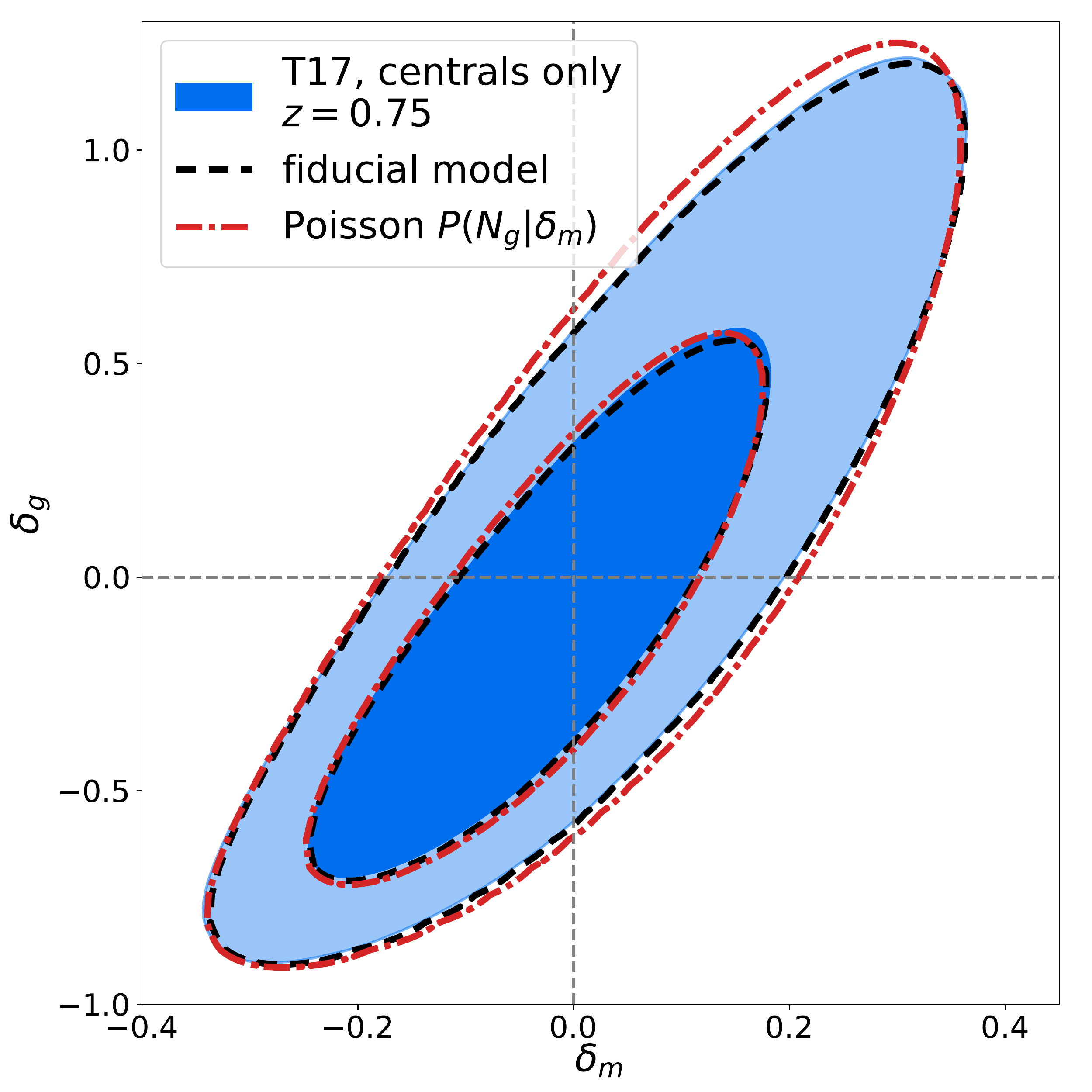}
  
  \includegraphics[width=0.4\textwidth]{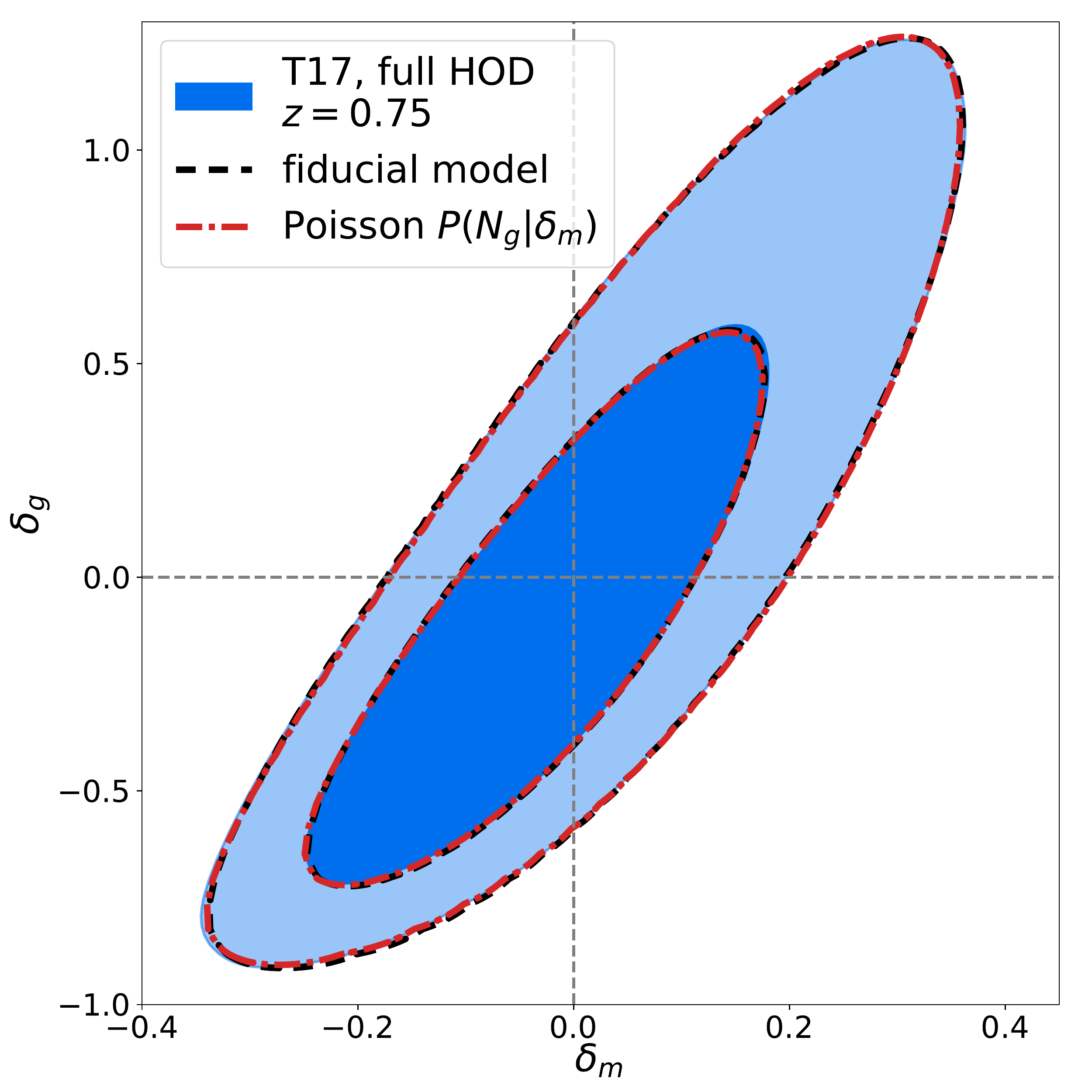}
  
  \includegraphics[width=0.4\textwidth]{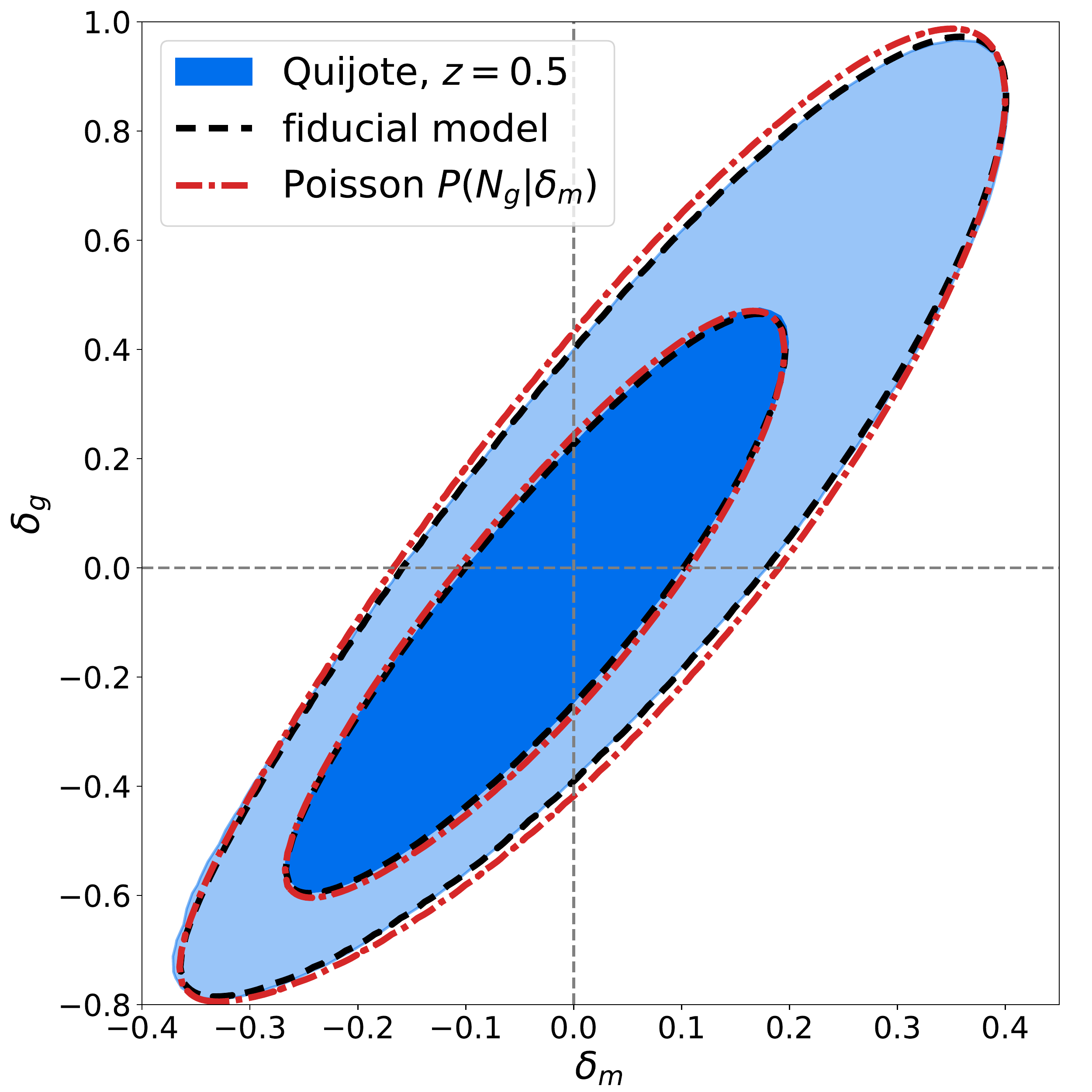}
  \end{center}
   \caption{Impact of non-Poisson shot-noise on the joint PDF of $\delta_g$ and $\delta_m$. The filled contours represent measurements in our mock data sets, the black dashed contours represent our fiducial model and the red dash-dotted contours show a model that assumes Poissonian shot-noise (see main text for details).}
  \label{fi:joint_PDFs_shot-noise}
\end{figure}

The upper panel of \figref{Quijote_vs_T17} shows the behaviour of shot-noise for the three different tracer samples in the Quijote data while the lower panel shows the measurements from the T17 data. One feature that persists in both data sets compared to what we found in \figref{shot-noise} is that satellite galaxies show an increase of $\mathrm{Var}(N_{\mathrm{tracer}})/\langle N_{\mathrm{tracer}} \rangle$ \wrt central galaxies that is almost independent of the total matter density $\delta_m$ in the smoothing aperture. But the shot-noise behaviour of the halo samples is quite different both between Quijote and T17 and compared to the $z\approx 0.75$ shell of T17. For all halo samples we considered there is a significant curvature of $\mathrm{Var}(N_{\mathrm{tracer}})/\langle N_{\mathrm{tracer}} \rangle$ as a function of $\delta_m$. But that curvature is strongest for the Quijote halos and even causes them to be super-Poissonian at very high densities. We could not find an obvious explanation for this difference but assume that it is caused by the different cosmologies at which the simulations are run (\cf \appref{mass_functions} for a comparison of the halo mass functions of the two simulations, and \figref{mass_functions} where it is shown that Quijote has significantly more high-mass halos). Given a precise measurement of the covariance $\mathrm{Cov}(N_i, N_j)$ of halo shot-noise in narrow mass bins as well as a model for the halo mass function and a given HOD we could in principle model $\mathrm{Var}(N|\delta_m)/\langle N|\delta_m\rangle$ exactly. There is, however, a number of practical reasons that prevent us from doing so:
\begin{itemize}
    \item As mentioned earlier in this section, measuring $\mathrm{Cov}(N_i, N_j)$ in sufficiently narrow bins will require a prohibitively large amount of simulations. Alternatively one could attempt to model the shot-noise covariance, but as of now no such model is available.
    \item HOD descriptions themselves make the assumption that satellite counts in a given halo are drawn from a Poisson distribution. This assumption is similarly adhoc as the assumption that galaxies are Poissonian tracers of the matter density field \citep[see \egnospace][who indeed find non-Poissonianity in the occupation distribution of sub-halos]{Boylan-Kolchin2010, Mao2015}.
    \item HOD descriptions also introduce a large number of free parameters which - unless they can be constrained a priori - may significantly dilute the cosmological constraining power of PDF analyses (or at least make them significantly more complicated).
\end{itemize}
Given our current (poor) understanding of shot-noise we hence conclude that effective parametrizations of non-Poisson shot-noise such as the one described in \secref{theory} are the most promising way forward for PDF analyses. In \figref{joint_PDFs_shot-noise} we show that this parametrization can indeed capture the impact of the non-Poissonianity observed in Figures~\ref{fi:shot-noise} and \ref{fi:Quijote_vs_T17} on the joint PDF of $\delta_m$ and $\delta_g$. The blue, filled contours in \figref{joint_PDFs_shot-noise} represent measurements of the PDF in our different mock data sets. For the black dashed lines we have fit a linear slope to our measurements of $\mathrm{Var}(N|\delta_m)/\langle N|\delta_m\rangle$ in order to determine the parameters $\alpha_0$ and $\alpha_1$ of our fiducial model presented in \secref{theory} (\cf the red dashed line in the upper panel of \figrefnospace{shot-noise} and analogous fits for the other tracer samples). The red dash-dotted contours represent an alternative model which assumes that shot-noise is exactly Poissonian. For the centrals in our fiducial T17 shell (\ie $z\approx 0.75$) such a model clearly overestimates the vertical width of the distribution (\cf upper panel). But for the full sample the values of $\mathrm{Var}(N|\delta_m)/\langle N|\delta_m\rangle$ become close to Poissonian again. Hence, even the Poissonian model accurately captures the shape of the joint PDF for that sample (middle panel). This is however only coincidental, and for our alternative sample in the Quijote simulations ($z=0.5$, full HOD) even the PDF of the full sample is noticeably different from the Poisson model (lower panel). The best-fitting parameters of our shot-noise model for the different galaxy samples we considered are summarized in \tabrefnospace{a0_a1_vals}.

Understanding shot-noise remains one of the most crucial tasks in the program of fully harvesting the information content of PDF-type analyses. Our results can serve as a foundation and starting point for that but they remain qualitative. We have also only qualitatively shown that our shot-noise model is effective in capturing the behaviour of the joint PDF $p(\delta_m, \delta_g)$ \citep[though see][for a quantitative analysis of the performance of this parametrization for density split statistics]{Friedrich2018}. To determine the accuracy of our model quantitatively, we need to specify a target survey (and hence target statistical uncertainties) as well as an observable that can replace matter density in our pair of $(\delta_m, \delta_g)$, since $\delta_m$ cannot directly be observed on real data. We leave this to the next step in our program and give a preliminary outlook in \secrefnospace{conclusions}.

\begin{table}
\begin{tabular}{|l|c|c|c}
tracer sample & redshift & $\alpha_0$ & $\alpha_1$ \\
\hline
T17 centrals & $0.751$ & $0.872\pm0.005$ & $-0.329\pm0.033$ \\
T17 all & $0.751$ & $1.015\pm0.006$ & $-0.310\pm0.043$ \\
T17 centrals & $0.476$ & $0.829\pm0.003$ & $-0.234\pm0.016$ \\
T17 all & $0.476$ & $0.945\pm0.003$ & $-0.274\pm0.018$ \\
Quijote centrals & $0.5$ & $0.608\pm0.001$ & $-0.061\pm0.004$ \\
Quijote all & $0.5$ & $0.868\pm0.001$ & $-0.142\pm0.007$ \\
\end{tabular}
    \caption{Best-fitting parameters of the shot-noise model from \secref{non_Poissonian_shot_noise_model} for the different galaxy samples considered in Figures~\ref{fi:Quijote_vs_T17} and \ref{fi:joint_PDFs_shot-noise}.}
\label{tab:a0_a1_vals}
\end{table}

\section{Discussion}
\label{sec:conclusions}

In this paper we investigated the relationship between the matter density field and its tracers from the PDF perspective, \ie the impact of the matter-tracer connection on the joint PDF $p(\delta_\mathrm{tracer},\delta_m)$. To evaluate this PDF we considered the matter density and tracer density fields in (long) cylindrical apertures as opposed to the spherical filters that are more commonly used in the theoretical literature. This choice was motivated by the fact that the cumulant generating function (CGF) of line-of-sight projected density fields can be expressed as Limber-type integral over CGFs of density fields in cylindrical apertures. Hence, it is only a small step to transfer our results to realistic observational situations of \eg photometric galaxy surveys.

The matter-tracer connection in the PDF context can be viewed as consisting of two ingredients: the conditional expectation value of $\delta_\mathrm{tracer}$ given $\delta_m$, $\langle \delta_\mathrm{tracer}|\delta_m\rangle$, and the scatter of $\delta_\mathrm{tracer}$ around this expectation value which is usually referred to as shot-noise. The fiducial model for $p(\delta_\mathrm{tracer},\delta_m)$ which we present here then consists of
\begin{itemize}
    \item a standard, large deviation theory (LDT) model for the PDF of matter density fluctuations $p(\delta_m)$, following the work of \eg \citet{Bernardeau1994, Bernardeau2000, Valageas2002, Friedrich2018, Uhlemann2018c, Barthelemy2019};
    \item a Lagrangian bias expansion for $\langle \delta_\mathrm{tracer}|\delta_m\rangle$, incorporated into the LDT formalism;
    \item a generalisation of the Poisson distribution as proposed by \citet{Friedrich2018, Gruen2018}.
\end{itemize}
Our Figures \ref{fi:joint_PDF} and \ref{fi:joint_PDFs_shot-noise} show that all of these aspects of our model are important for describing the full shape of $p(\delta_\mathrm{tracer},\delta_m)$. In the following two subsections we first summarize the results of our study and then briefly discuss open tasks for PDF cosmology.

\subsection{Summary of results}

We have added a number of tools and observations to the already rich subject of cosmic density PDFs:
\begin{itemize}
    \item We consistently incorporated a Lagrangian bias expansion for the conditional expectation value $\langle \delta_{\mathrm{tracer}}|\delta_m \rangle$ into the standard LDT formalism for modelling cosmic density PDFs. We also demonstrated that at the saddle point configuration of the initial density field which determines the LDT predictions (\cf the path integral in \eqnrefnospace{Friedrich_functional_integral}) the operations of filtering and squaring the density field approximately commute (\cf \figrefnospace{commutation}). This makes it possible to evaluate the Lagrangian expansion up to 2nd order with essentially no additional computational coast. An advantage of our Lagrangian model that we did not discuss here is that it allows one to consistently incorporate scale-dependent bias from primordial non-Gaussianity \citep[\egnospace][]{Dalal2008, Desjacques2009, Jeong2009} into the LDT formalism. This is because our ansatz in \secref{Lagrangian_model} can be used to translate scale-dependence of $b_1^L$ into a density dependence.
    \item We fitted both the Langrangian and an Eulerian expansion to measurements of $\langle \delta_{\mathrm{halo}}|\delta_m \rangle$ for different halo mass bins and at different redshifts and filtering scales in simulated data by \citet{Takahashi2017}. In this way we could validate that the bias expansion we developed in \secref{theory} conforms to standard consistency relations between the Eulerian and Lagrangian perspective of halo bias. We also checked for the consistency of our best-fitting bias parameters with expectations from other methods: our values of $b_1^L$ as a function of halo mass agree well with a number of different theoretical and empirical predictions; the relation we observe between $b_1^L$ and $b_2^L$ agrees with an empirical formula found by \citet{Lazeyras2016} in separate universe simulations; and for large smoothing scales our best-fitting linear bias converges to the corresponding parameter measured from the large-scale cross power spectrum of matter and galaxy density. This array of tests confirms that the theory we developed in \secref{theory} represents a sensible Lagrangian bias model and hence moves PDF analyses one step closer to being on equal footing with the more advanced field of N-point correlation functions. We also showed that for a synthetic galaxy sample mimicking eBOSS-like luminous red galaxies, the Lagrangian expansion yields a significantly better fit to $\langle \delta_{\mathrm{halo}}|\delta_m \rangle$ than the Eulerian expansion at second order. This is however not a general statement and we saw indications that for very massive halos the Eulerian expansion performs better.
    \item We established that the deviation of shot-noise from Poisson noise in a sample of halos with a wide mass range is determined by the covariance matrix $\mathrm{Cov}(N_i, N_j)$ of the shot-noise of halos in a very narrow binning of that mass range. Considering the ratio $\mathrm{Var}(N_\mathrm{tracer})/\langle N_\mathrm{tracer} \rangle$ for different tracer samples in both the Quijote and T17 simulations we have found a wide variety of deviations from Poissonian shot-noise. We have however shown that our shot-noise model from \secref{non_Poissonian_shot_noise_model} is effective in capturing the impact of these deviations on the joint PDF $p(\delta_g,\delta_m)$.
\end{itemize}

As mentioned in the previous section, our results on shot-noise remain qualitative and more insights may be needed to efficiently model that part of the PDF. We discuss this further in the following outline.

\subsection{Open tasks for PDF cosmology}

Cosmological analyses of the full shape of $p(\delta_\mathrm{tracer},\delta_m)$ can be seen as an extension of the density split statistics framework developed by \citet{Friedrich2018, Gruen2018}. In year-1 data of the Dark Energy Survey (DES) they have analysed a data vector consisting of (a compressed version of) the galaxy density PDF $p(\delta_\mathrm{tracer})$ and a number of lensing signals, that effectively probe the expectation values $\langle \delta_m | \delta_\mathrm{tracer} \rangle$ as well as the slope of the lensing power spectrum. Moving away from these compressed statistics and directly analysing $p(\delta_\mathrm{tracer},\delta_m)$ instead will, at any given smoothing scale, open up an entire 2-dimensional plane of data for cosmological analysis. There is a number of steps that still need to be completed to implement this program.

\textbf{Cosmological constraining power:} Numerous studies have shown that the cosmological information contained in the PDF of density fluctuations strongly complements the information obtained from more standard probes such as the 2-point correlations of fluctuations - see \eg \citet{Codis2016a, Patton2017, Uhlemann2020, Friedrich2020, Boyle2021} for recent examples. Some of their results however only apply to idealised situations where one has direct access to the matter density field and the question remains to what extent the cosmological power of the PDF carries over to realistic data sets. \citet{Boyle2021} have considered the PDF of lensing convergence, which can in principle be obtained from observations of cosmic shear. And \citet{Friedrich2018, Gruen2018} analysed compressed statistics of the joint PDF $p(\delta_\mathrm{tracer},\delta_m)$, showing that it has a competitive power to constrain cosmological models. But as mentioned above, their density split statistics are also sensitive to the slope of the lensing power spectrum, and that information would be lost if one would only consider $p(\delta_\mathrm{tracer},\delta_m)$ at one smoothing scale. Two solutions to this problem would be to analyse the PDF at a number of different smoothing scales \citep[as was \eg done by][]{Boyle2021} or to analyse the joint PDF of galaxy densities in apertures that are located at a finite distance \citep[a 2-point PDF, \cfnospace][]{Uhlemann2018a}. Alternatively, one could consider combined analyses of the PDF and the 2-point function. We have shown that at large scales the linear bias of a PDF analysis agrees with the large scale bias of the tracer-matter cross power spectrum. This would suggest that a combination of a PDF-type analysis with measurements of the galaxy-galaxy lensing correlation function \citep[gg-lensing, see \egnospace][and references therein]{Prat2017} is a promising route to take. To efficiently analyse such a combined data vector, one will need to make contact between the shot-noise and higher order bias parameters of our PDF model and stochasticity effects and non-linear biasing in the gg-lensing correlation function. This leads us to the next point.

\textbf{Improved modelling:} In the model presented here, the galaxy-matter connection is described by 4 free parameters. While \citet{Friedrich2018, Gruen2018} have shown that the rich information content of the PDF can constrain complex bias models, a more efficient modelling would be highly desirable. This can \eg be achieved by choosing informative, physically motivated priors on our parameters (\cf Britt et al., Ried et al.\ in prep.) or by identifying consistency relations between them. For example, non-linear bias at a small scale will lead to an effective change in shot-noise at a larger scale \citep{Philcox2020}, which should lead to a relation between bias and the scale dependence of shot-noise. Understanding these kinds of relations will also enable a more fruitful combination of PDF and 2-point function analyses, and the information present in the PDF may be able to constrain nuisance parameters in 2-point function models.

\textbf{Proof of concept:} A more immediate goal which we envision as a follow-up to this study is a proof-of-concept study that demonstrates the feasibility of analysing the full shape of $p(\delta_\mathrm{tracer},\delta_m)$ in real data. Since matter density is not directly directly observable, we aim at the joint PDF of lensing convergence and 2D-projected galaxy density. The results of this paper can be readily generalised to such line-of-sight projected fields \citep[see \egnospace][]{Bernardeau2000, Friedrich2018, Uhlemann2018c, Barthelemy2019, Boyle2021}, so such an analysis is indeed within reach.


\section*{Acknowledgements} 

We would like to thank Bernardita Ried, Zvonimir Vlah and Risa Wechsler for helpful discussions.

OF gratefully acknowledges support by the Kavli Foundation and the International Newton Trust through a Newton-Kavli-Junior Fellowship and by Churchill College Cambridge through a postdoctoral By-Fellowship.
SC's work is partially supported by the SPHERES grant ANR-18-CE31-0009 of the French {\sl Agence Nationale de la Recherche} and by Fondation MERAC.
This work was supported by the Department of Energy, Laboratory Directed Research and Development program at SLAC National Accelerator Laboratory, under contract DE-AC02-76SF00515 and as part of the Panofsky Fellowship awarded to DG.

\section*{Data availability} 

\verb|C++| and \verb|python| tools to compute our model predictions are publicly available at \url{https://github.com/OliverFHD/CosMomentum} . The data for the T17 N-body simulations used in this article are publicly available at \url{http://cosmo.phys.hirosaki-u.ac.jp/takahasi/allsky_raytracing/} . Summary statistics measured in the Quijote N-body simulations are publicly available at \url{https://github.com/franciscovillaescusa/Quijote-simulations}. The Molino mock galaxy catalogs are publicly available at \url{https://changhoonhahn.github.io/molino/current/} .

\def\aj{AJ}%
\def\araa{ARA\&A}%
\def\apj{ApJ}%
\def\apjl{ApJ}%
\def\apjs{ApJS}%
\def\ao{Appl.~Opt.}%
\def\apss{Ap\&SS}%
\def\aap{A\&A}%
\def\aapr{A\&A~Rev.}%
\def\aaps{A\&AS}%
\def\azh{AZh}%
\def\baas{BAAS}%
\def\jrasc{JRASC}%
\def\memras{MmRAS}%
\def\mnras{MNRAS}%
\def\pra{Phys.~Rev.~A}%
\def\prb{Phys.~Rev.~B}%
\def\prc{Phys.~Rev.~C}%
\def\prd{Phys.~Rev.~D}%
\def\pre{Phys.~Rev.~E}%
\def\prl{Phys.~Rev.~Lett.}%
\def\pasp{PASP}%
\def\pasj{PASJ}%
\def\qjras{QJRAS}%
\def\skytel{S\&T}%
\def\solphys{Sol.~Phys.}%
\def\sovast{Soviet~Ast.}%
\def\ssr{Space~Sci.~Rev.}%
\def\zap{ZAp}%
\def\nat{Nature}%
\def\iaucirc{IAU~Circ.}%
\def\aplett{Astrophys.~Lett.}%
\def\apspr{Astrophys.~Space~Phys.~Res.}%
\def\bain{Bull.~Astron.~Inst.~Netherlands}%
\def\fcp{Fund.~Cosmic~Phys.}%
\def\gca{Geochim.~Cosmochim.~Acta}%
\def\grl{Geophys.~Res.~Lett.}%
\def\jcap{JCAP}%
\def\jcp{J.~Chem.~Phys.}%
\def\jgr{J.~Geophys.~Res.}%
\def\jqsrt{J.~Quant.~Spec.~Radiat.~Transf.}%
\def\memsai{Mem.~Soc.~Astron.~Italiana}%
\def\nphysa{Nucl.~Phys.~A}%
\def\physrep{Phys.~Rep.}%
\def\physscr{Phys.~Scr}%
\def\planss{Planet.~Space~Sci.}%
\def\procspie{Proc.~SPIE}%

\bibliographystyle{mnras}
\bibliography{literature}

\appendix

\section{Equations of motion for cylindrical collapse}
\label{app:cylindrical_collapse}

We repeat here an appendix of \citet{Friedrich2020} about the evolution of symmetric density perturbations. In the Newtonian approximation and setting $G = 1 = c$ the evolution of spherical, cylindrical or planar perturbations $\delta$ is described by
\begin{equation}
\label{eq:SC_in_conformal_time}
\ddot{\delta} + \mathcal{H} \dot{\delta} - \frac{N+1}{N} \frac{\dot{\delta}^2}{1+\delta} \ = 4\pi \bar \rho_m a^2 \delta (1+\delta)\ ,
\end{equation}
where $\tau$ is conformal time, $\mathcal{H} = \dd \ln a / \dd \tau$ is the conformal expansion rate and $N=3$ for a spherical perturbation, $N=2$ for a cylindrical perturlation and $N=1$ for a planar perturbation (see \citealt{MukhanovBook} who demonstrates this for $N=1$ and $N=3$). To compute the evolution of the saddle point fluctuation in \secref{cylinder_CGF_with_bias} we choose $N=2$ and solve \eqnref{SC_in_conformal_time} with the initial conditions
\begin{equation}
    \delta_i = \delta_{\mathrm{lin},R_{\mathrm{lin}}}^*\ D(z_i)\ ,\ \dot{\delta_i} = \delta_i\ \mathcal{H}(z_i)\ ,
\end{equation}
where $z_i$ is a redshift chosen during matter domination. (In fact, in our calculation of $D(z)$ we set the radiation density $\Omega_r$ to zero and then choose $z_i = 4000$.)

\section{Cylindrical average of the squared linear saddle point}
\label{app:commutation}

Adjusting the results of \citet{Valageas2002, Friedrich2020} to cylindrical filters, the saddle point configuration of the linear density contrast, $\delta_{\mathrm{lin}}^*$, filtered with a cylindrical aperture of radius $R$ is given by
\begin{equation}
    \delta_{\mathrm{lin}, R}^* = \delta_{\mathrm{lin}, R_{\mathrm{lin}}}^*\ \frac{\langle \delta_{\mathrm{lin}, R}\ \delta_{\mathrm{lin}, R_{\mathrm{lin}}}\rangle}{\langle \delta_{\mathrm{lin}, R_{\mathrm{lin}}}^2\rangle}\ .
\end{equation}
Here we have assumed Gaussian initial conditions \citep[see][for general non-Gaussian initial conditions]{Friedrich2020b} and we have set $\lambda_h = 0$, which is the case that is of interest for the calculation of $\langle \delta_{\mathrm{halo}}|\delta_m\rangle$ (\cf \eqnrefnospace{conditional_expectation_value}).

At any point $\bm{r}$ the saddle point configuration is then given by
\begin{equation}
    \delta_{\mathrm{lin}}^*(\bm{r}) = \delta_{\mathrm{lin}, r}^* + \frac{r}{2} \left.\frac{\dd \delta_{\mathrm{lin}, R'}^*}{\dd R'}\right|_{R' = r}\ .
\end{equation}
In \eqnref{bias_at_saddle_point} we need to know the average of $\delta_{\mathrm{lin}}^*(\bm{r})^2$ in cylindrical apertures. This average can be calculated as
\begin{equation}
    [{\delta_{\mathrm{lin}}^*}^2]_{R_{\mathrm{lin}}} = \frac{2}{R_{\mathrm{lin}}^2}\int_0^{R_{\mathrm{lin}}} \dd r\ r\ \left\lbrace \delta_{\mathrm{lin}, r}^* + \frac{r}{2} \left.\frac{\dd \delta_{\mathrm{lin}, R'}^*}{\dd R'}\right|_{R' = r} \right\rbrace^2\ .
\end{equation}
In \figref{commutation} we show that on the scales we are interested in, this full computation is well approximated by simply squaring the cylindrically averaged saddle point configuration. This approximation will bias our values of quadratic Lagrangian bias by a couple of percent \wrt other measures of bias, which does not significantly affect the conclusions of our study.

\section{Comparing the mass functions of Quijote and T17}
\label{app:mass_functions}

\begin{figure}
\begin{center}
  \includegraphics[width=0.48\textwidth]{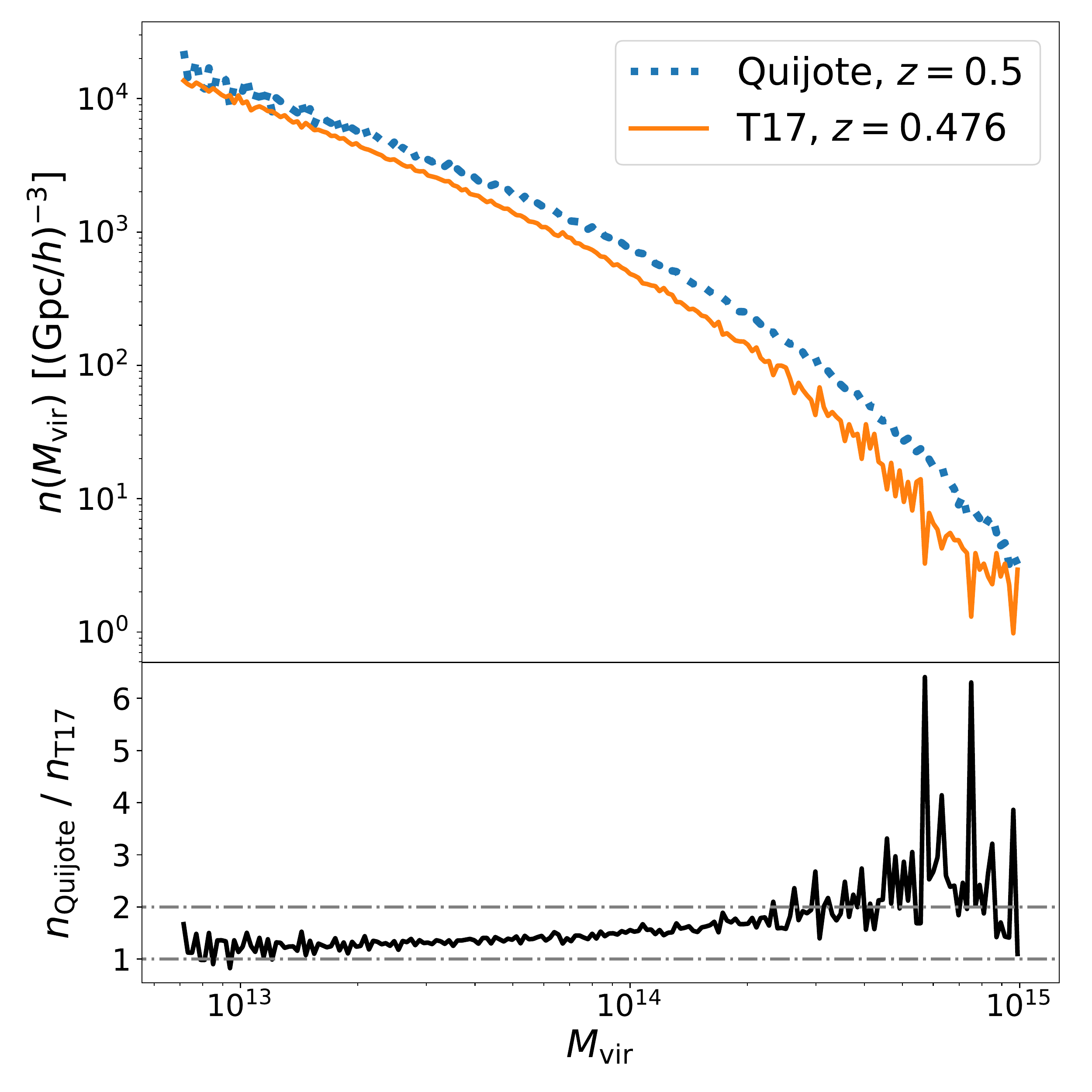}
  \end{center}
   \caption{Upper panel: the mass function $n(M_{\mathrm{vir}})$ of the two different N-body data sets considered in \secref{shot-noise_details} at $z=0.5$ (Quijote) and $z=0.476$ (T17). We are only plotting $n(M_{\mathrm{vir}})$ above the mass cut of $M_{\mathrm{vir}}=6.986\cdot 10^{12} M_\odot/h$ that we considered in that section. Lower panel: ratio of the mass functions in the two simulations.}
  \label{fi:mass_functions}
\end{figure}

In \figref{mass_functions} we compare the mass function $n(M_{\mathrm{vir}})$ of the two different N-body data sets considered in \secref{shot-noise_details} at $z=0.5$ (Quijote) and $z=0.476$ (T17). Our reason for using $M_{\mathrm{vir}}$ is that we do not have $M_{200\mathrm{b}}$ available for Quijote. The differences in the mass functions are likely caused by the different cosmology of the simulations - $(\Omega_m, \Omega_b, \sigma_8, n_s, h) = (0.3175, 0.049, 0.834, 0.9624, 0.6711)$ for Quijote and $(0.279, 0.046, 0.82, 0.97, 0.7)$ for T17. We think that this difference in cosmology and the mass function is at least in part responsible for the differences in shot-noise behaviour of the two data sets that we observed in \secrefnospace{shot-noise_details}.

\end{document}